\documentclass[lettersize,journal]{IEEEtran}

\usepackage{times}
\usepackage{latexsym}

\usepackage[T1]{fontenc}

\usepackage[utf8]{inputenc}
\usepackage[numbers]{natbib}

\usepackage{microtype}

\usepackage{inconsolata}

\usepackage{graphicx}
\usepackage{enumitem}
\usepackage{booktabs}
\usepackage{tabularx}
\usepackage{amsmath} 
\usepackage{xspace}
\usepackage{tcolorbox}
\usepackage{xcolor}
\definecolor{grey}{gray}{0.9}
\usepackage{multirow} 
\usepackage{listings}
\usepackage{color}
\definecolor{dkgreen}{rgb}{0,0.6,0}
\definecolor{gray}{rgb}{0.5,0.5,0.5}
\definecolor{mauve}{rgb}{0.58,0,0.82}
\newcommand{\dataset}{CoSQA\textsuperscript{+}\xspace}
\newcommand{\all}{CoSQA\textsuperscript{+}\_all\xspace}
\newcommand{\verified}{CoSQA\textsuperscript{+}\_verified\xspace}

\newcommand{\figmargin}{\vspace{-4pt}}
\newcommand{\tabmargin}{\vspace{-4pt}}

\newcommand{\boxmargin}{2mm}
\tcbuselibrary{most, skins, breakable, theorems}
\newtcolorbox{myboxa}[2][]{
    colback=gray!10!white,
    colframe=black, enhanced,
    attach boxed title to top left={yshift=-2mm,xshift=5mm},
    title=#2,#1
}
\newtcolorbox{myboxb}[2][]{
    boxsep=3pt,
    left = \boxmargin, right = \boxmargin, top = \boxmargin, bottom = \boxmargin,
    title={#2},#1
}
\newtcolorbox{myboxc}{
    colback=gray!15!white,
    arc = 0pt, outer arc = 0pt,
    boxsep=0pt, left = 3pt, right = 0pt, top = 0pt, bottom = 0pt, 
    leftrule=3pt, bottomrule=0pt,toprule=0pt, rightrule=0pt,
    left = \boxmargin, right = \boxmargin, top = \boxmargin, bottom = \boxmargin
}
\newtcolorbox{myboxd}{
    colback=gray!10,
    colframe=black,
    width=\columnwidth,
    arc=1mm, auto outer arc,
    boxrule=0.5pt,
}

\usepackage{times}  
\usepackage{helvet}  
\usepackage{courier}  
\usepackage{graphicx} 
\usepackage{hyperref} 
\usepackage{cleveref} 
\usepackage{natbib}  
\usepackage{caption} 
\usepackage{algorithm}
\usepackage{algorithmic}
\newcommand{\revised}[1]{\textcolor{black}{#1}}
\newcommand{\MRrevised}[1]{\textcolor{black}{#1}}
\usepackage{newfloat}
\usepackage{listings}
\DeclareCaptionStyle{ruled}{labelfont=normalfont,labelsep=colon,strut=off} 
\floatstyle{ruled}
\newfloat{listing}{tb}{lst}{}
\floatname{listing}{Listing}
\usepackage{listings}
\usepackage{array}
\usepackage{longtable} 

\usepackage{amsmath,amssymb,amsfonts}
\usepackage{placeins}
\usepackage{needspace}
\usepackage{afterpage}
\begin{document}

\title{\dataset: Enhancing Code Search Evaluation with a Multi-Choice Benchmark and Test-Driven Agents}

\author{Jing Gong\IEEEauthorrefmark{1}, Yanghui Wu\IEEEauthorrefmark{1}, Linxi Liang\IEEEauthorrefmark{1}, Yanlin Wang\IEEEauthorrefmark{2}, Jiachi Chen, Mingwei Liu, and Zibin Zheng~\IEEEmembership{Fellow,~IEEE}\thanks{Jing Gong, Yanghui Wu, Linxi Liang, Yanlin Wang, Jiachi Chen, Mingwei Liu, Zibin Zheng are with the School of Software Engineering, Sun Yat-sen University, Guangzhou, Guangdong 510275, China (e-mail: gongjing13424@gmail.com; thinkerhui@qq.com; lianglx26@mail2.sysu.edu.cn; yanlin-wang@outlook.com; chenjch86@mail.sysu.edu.cn, liumw26@mail.sysu.edu.cn, zhzibin@mail.sysu.edu.cn )}}

\maketitle

\begin{abstract}

Semantic code search, retrieving code that matches a given natural language query, is an important task to improve productivity in software engineering.  
Existing code search datasets face limitations: they rely on human annotators who assess code primarily through semantic understanding rather than functional verification, leading to potential inaccuracies and scalability issues. Additionally, current evaluation metrics often overlook the multi-choice nature of code search.
This paper introduces \dataset, pairing high-quality queries from CoSQA with multiple suitable codes. 
We develop an automated pipeline featuring multiple model-based candidate selections and the novel test-driven agent annotation system. Among a single Large Language Model (LLM) annotator and Python expert annotators (without test-based verification), agents leverage test-based verification and achieve the highest accuracy of \revised{93.9\%}. Through extensive experiments, \dataset has demonstrated superior quality over CoSQA. Models trained on \dataset exhibit improved performance. \revised{We publicly release both \all, which contains 412,080 agent-annotated pairs, and \verified, which contains 1,000 human-verified pairs, at \url{https://github.com/DeepSoftwareAnalytics/CoSQA_Plus}.}

\end{abstract}


\renewcommand{\thefootnote}{\IEEEauthorrefmark{1}} 
\footnotetext{Equal Contribution.}
\renewcommand{\thefootnote}{\arabic{footnote}} 

\renewcommand{\thefootnote}{\IEEEauthorrefmark{2}} 
\footnotetext{Corresponding Author.}
\renewcommand{\thefootnote}{\arabic{footnote}} 

\section{Introduction}

\IEEEPARstart{S}{emantic} code search is to retrieve codes that match a given natural language query, which is important for accelerating software development~\cite{li2013help, sim2011well, stolee2014solving, panichella2013effectively, uchitel2024scoping}. Programmers can describe their needs in natural language, such as implementing specific algorithms, using particular APIs, or solving complex problems~\cite{yan2020code}. \revised{For the given query, while existing semantic code search studies focus on retrieving relevant codes, we define a different task: to retrieve functionally matching codes that satisfy the query’s requirements when executed in a complete context.}

\revised{Although one query can be matched by multiple codes,} existing code search benchmarks are designed for one-to-one query-code matching tasks and are typically evaluated using Mean Reciprocal Rank (MRR)~\cite{husain2019codesearchnet, huang2021cosqa,li2024optimizing,lu2021codexglue,yao2018staqc, feng2020codebert}. \revised{This design leads to a critical mismatch with real-world practices.} According to our survey of 200 experienced Python programmers with an average of over 1 year of experience, they perform code searches approximately 7.79 times per programming day on average. For each query, they typically reference \revised{2.83} code examples to find the necessary information.
\revised{Furthermore, participants report that 63.2\% of their queries commonly result in multiple valid code snippets. This reveals that developers operate in a multi-choice context, yet existing benchmarks adhere to a single-choice evaluation paradigm with only one "correct" code snippet per query.} 




\revised{The construction of benchmarks that reflect this multi-choice reality faces significant challenges, as methodologies designed for the prevalent one-to-one task (e.g., CoSQA~\cite{huang2021cosqa}, CodeSearchNet~\cite{husain2019codesearchnet}) are hampered by issues of scalability, accuracy, and evaluation:}


\begin{enumerate}[label=C\bfseries\arabic*, leftmargin=*, itemsep=0pt, parsep=0pt, partopsep=0pt, topsep=0pt]

\item \textbf{Scalability Issues in Human-Based Benchmarking.} Conventional benchmarks like CoSQA~\cite{huang2021cosqa} and CodeSearchNet (CSN)~\cite{husain2019codesearchnet} rely on expert human annotators to evaluate query-code pairs. While this approach captures human judgment, manual annotation requires significant time and resources, limiting dataset scalability. The labor-intensive nature of this process makes it impractical for creating large-scale benchmarks needed for robust code search evaluation.

\item \textbf{Accuracy Limitations in Annotation Methods.} Both human and LLM-based annotation approaches suffer from accuracy issues. \revised{Human annotators typically assess code matching based on surface-level comprehension rather than verifying functional correctness~\cite{dunlop1982comparative} through test execution, which may lead to inaccurate annotations due to a lack of full understanding of syntax, semantics, and behavioral correctness.} While Li et al. leverage LLMs to address scalability issues~\cite{li2024optimizing}, these models struggle to reliably determine whether code functionally satisfies query requirements, as they tend to focus on semantic similarity rather than execution logic. This fundamental limitation affects the quality of both human and LLM-based annotations~\cite{liu2023your}.

\item \revised{\textbf{Lack of Evaluation in Multi-Choice Scenarios.} One-to-one Conventional one-to-one matching benchmarks typically rely on MRR, which measures the rank position of the first relevant code snippet. However, in practical scenarios, developers often seek multiple code examples for reference, where the key concern is how many relevant snippets appear within the top-10 results or on the first page. MRR and other single-choice benchmarking approaches fail to capture this dimension of performance.}


\end{enumerate}



To address these limitations, we present a new benchmark \textbf{\dataset}
\revised{, comprising two releases: \all with 412,080 pairs of web queries from CoSQA and repository code curated from CodeSearchNet, and \verified with 1,000 human-verified pairs that serve as a gold-standard evaluation subset.}
\revised{The benchmark is specifically designed for multi-choice code search and adopts MAP as the primary evaluation metric.}

To construct high-quality pairs and achieve a high coverage rate of positive pairs, we employ multiple models to select the top 20 codes with the highest similarity for each query, followed by our novel test-driven agents - a fully automated pipeline comprising a preliminary screener, a test program generator, a test executor, a bug fixer, and a final arbiter. 
The label indicates whether each code snippet exactly matches its corresponding query requirements. \revised{Specifically, exactly matched codes refer to codes whose functionality fully satisfies or exceeds the requirements specified in the query. Furthermore, we evaluate 8 code search methods—7 of which are embedding-based.}


To assess the efficacy of our proposed annotation approach, we establish a rigorous evaluation framework using 1000 randomly selected query-code pairs from \dataset. Three Python experts first design test programs for these pairs to establish ground truth annotations. 
We then compare the accuracies among test-driven agents, python expert annotators without tests, and a single LLM. Results demonstrate that the test-driven agents achieve superior performance with an accuracy of \revised{93.9\%}. Notably, 83.67\% of the test programs automatically generated by test-driven agents are executable. 
 Furthermore, using identical data sources (queries from CoSQA and code from CodeSearchNet), we evaluate \dataset against CoSQA as fine-tuning corpora for three models: CodeBERT\footnote{The version of CodeBERT used in this paper is fine-tuned on the code search task with CodeSearchNet, not the original version.}, UniXcoder, and CodeT5+ embeddings.
 Across all three models, fine-tuning on \dataset consistently yields superior performance on the CSN99 Python benchmark, as measured by both \revised{MAP(Mean Average Precision)@10} and MRR metrics. \revised{Furthermore, we demonstrate that the test-driven agent generalizes beyond Python, achieving consistent accuracy gains of 10.33\% on average across PHP, Java, and Go (RQ5).}



Our contributions can be summarized as follows:
\begin{itemize}
\item To the best of our knowledge, we develop the first Multiagent annotator with \revised{93.9\%} accuracy for code search dataset.
\item To the best of our knowledge, we construct \dataset which is the first benchmark designed for evaluating multi-choice code search. 
\item We enhance the fully automated construction process through candidate pair construction by multiple models and annotation by agents.
\item \revised{We publicly release both \all and \verified to facilitate further research in semantic code search.}
\end{itemize}

\section{Motivation Survey}

\begin{figure*}
    \centering
    \includegraphics[width=1.0\linewidth]{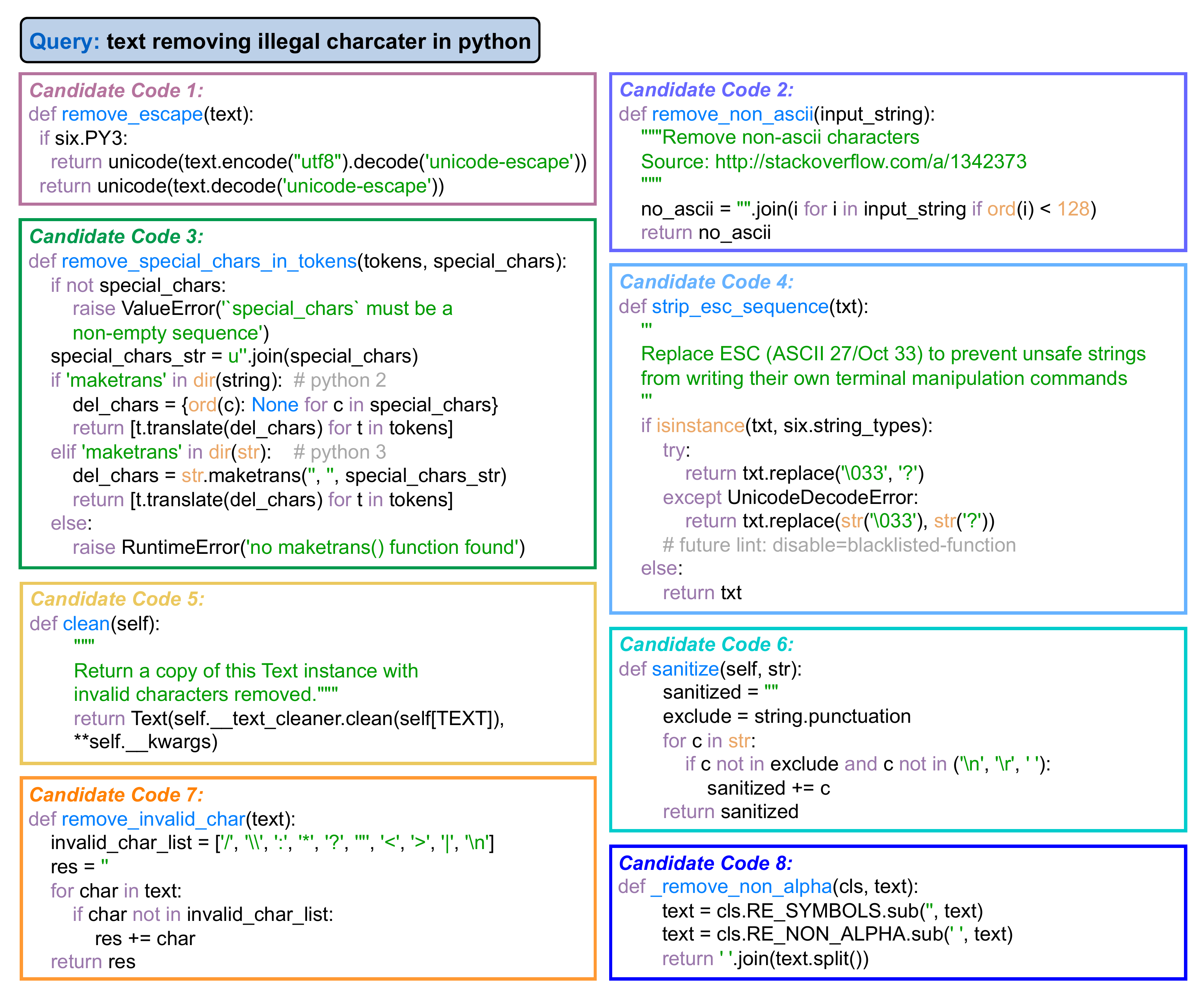}
    \caption{An example of multi-choice code search. The query is vague in its definition of "illegal character," allowing eight candidate code snippets to match under different interpretations.}
    \label{fig:motivation}
\end{figure*}

\textit{Do developers really need multi-choice code search?} To answer this question, we conduct a survey to investigate the code search behaviors and preferences of Python developers, with a particular focus on understanding how developers interact with multiple code examples during their search process. Fig.~\ref{fig:motivation} provides a multi-choice code search example. \revised{This example illustrates a common real-world scenario where query ambiguity leads to multiple legitimate interpretations. The term "illegal character" can reasonably refer to filesystem-invalid characters, non-ASCII characters, escape sequences, or special symbols, each representing valid use cases in different programming contexts.}

\paragraph{Methodology}
\revised{We recruited participants through online anonymous questionnaire platforms, targeting two main groups: current university students and working technical personnel in computer science or related fields. The survey collected responses from 200 participants, including 148 current students and 52 working professionals. In terms of educational background, the majority of participants hold bachelor's degrees (87.5\%) with a smaller proportion holding master's or doctoral degrees (11.5\%). Their Python programming experience varies, with the majority (59.0\%) having between 1 and 5 years of programming experience, while 16.0\% have between 5 and 10 years of experience. Among all participants, 64.0\% use Python as their primary programming language.}
No personally identifiable information is collected, and participants are informed that their responses will be used for research purposes only. 

\paragraph{Results and Analysis}
Our survey reveals several significant patterns in code search behavior \revised{across different user groups. Importantly, the need for multiple code examples is not limited to undergraduate students but extends to experienced developers as well. We find that 75.0\% of working technical professionals typically require 2-3 code examples to understand and implement a solution, compared to 63.5\% of current students. This finding challenges the assumption that multi-choice code search primarily serves educational purposes for beginners. In fact, experienced developers demonstrate an even higher tendency to reference multiple implementations, particularly when dealing with complex or unfamiliar tasks. Additionally, the participants reported that 63.2\% of their queries commonly result in multiple code snippets that meet their requirements, indicating that query ambiguity leading to multiple valid solutions is a prevalent phenomenon in real-world development scenarios. Code searchers prefer function codes with clear input and output (85.5\%), indicating a strong preference for code examples that explicitly define parameters and expected results regardless of their experience level. Participants evaluate search results based on several key factors, primarily focusing on solution differences (66.5\%) and implementation relevance (66.5\%).} These preferences emphasize that developers do not just want the closest match to their query, but also value diverse approaches that help them understand the problem and solution better. When presented with multiple code examples, they particularly value the clarity of implementation differences and the completeness of edge case handling.

\section{Related Work}

\subsection{Code Search Datasets}

Besides our work, there are other text-to-code datasets and benchmarks in the realm of code search. They can be categorized into three types based on the origin of the queries: doc-code datasets, question-code datasets, and query-code datasets.

CodeSearchNet~\cite{husain2019codesearchnet}, CoDesc~\cite{hasan2021codesc}, PyMT5~\cite{clement-etal-2020-pymt5}, and XLCoST~\cite{zhu2022xlcost} are doc-code datasets, which contain massive code snippets from public repositories with corresponding comments. They lay foundational training grounds for code search models to grasp basic code-comment relationships. CodeSearchNet~\cite{husain2019codesearchnet} dataset comprises a large-scale collection of query-code pairs extracted from GitHub repositories. CoDesc~\cite{hasan2021codesc} is a comprehensive dataset featuring 4.2 million Java methods paired with natural language descriptions, meticulously filtered to minimize noise. PyMT5, on the other hand, offers a vast collection of 26 million Python methods and 7.7 million method-docstring pairs, enabling bidirectional translation between code and docstrings. XLCoST~\cite{zhu2022xlcost} stands out as the largest multilingual parallel dataset, encompassing 8 languages. It supports 10 cross-lingual code tasks, such as translation and search, and provides essential baselines for model development in diverse programming contexts.

StaQC~\cite{yao2018staqc}, CoNaLa~\cite{yin2018mining}, SO-DS~\cite{heyman2020neural}, CS1QA~\cite{lee2022cs1qa}, and ProCQA~\cite{li2024procqa} are question-code datasets, which contain code answers with corresponding questions from Stack Overflow. They challenge models to excel in practical problem-solving scenarios by understanding complex developer inquiries.
StaQC~\cite{yao2018staqc} is a large dataset automatically mined from Stack Overflow, which includes multiple languages but varies in code quality. CoNaLa~\cite{yin2018mining} is a curated dataset derived from Stack Overflow, comprising 2,379 training and 500 test examples of natural language-code pairs, alongside a large automatically-mined dataset of 600k examples, designed for tasks like code synthesis, retrieval, and summarization. SO-DS~\cite{heyman2020neural}, a high-quality snippet collection mined from highly upvoted Stack Overflow posts, focuses on annotated code search, offering 12,137 snippets and 7,674 unique descriptions, with 2,225 annotated queries for validation and testing, specifically tailored for improving code retrieval systems by leveraging natural language descriptions. CS1QA~\cite{lee2022cs1qa}, a dataset for code-based question answering in programming education, comprises 9,237 annotated question-answer pairs from Python chat logs, alongside 17,698 unannotated examples, designed to benchmark models on code comprehension and question answering in an educational context. ProCQA~\cite{li2024procqa} introduces a large-scale community-based programming question answering dataset, which is mined from StackOverflow with strict filtering strategies for quality and fairness control. This benchmark adopts the code-mixing dataset, which is more aligned with real-world scenarios. However, the data source is still limited, and the criteria for judging whether the answer matches the question is still vague.

CoSQA~\cite{huang2021cosqa}, CSN99 of CodeSearchNet~\cite{husain2019codesearchnet}, WebQueryTest of CodeXGLUE~\cite{lu2021codexglue}, and Query4Code~\cite{li2024optimizing} are query-code datasets, which contain real human queries from web and code snippets matched with them. They are crucial for training and testing models' capabilities in handling authentic, diverse search queries. CoSQA~\cite{huang2021cosqa} introduces a dataset of 20,604 human-annotated natural language query-code pairs, enhancing semantic matching for code search, and proposes CoCLR, a contrastive learning method that boosts code question answering accuracy by 10.5\% when combined with CodeBERT. CSN99, part of the CodeSearchNet Challenge~\cite{husain2019codesearchnet}, offers 99 natural language queries with approximately 4k expert-annotated relevance judgments, facilitating robust evaluation of code search models. The queries of CoSQA and CSN99 are collected from Microsoft Bing search engine. WebQueryTest~\cite{lu2021codexglue} addresses the gap between real user search queries and existing code search datasets by providing a testing set of about 1k real web queries paired with Python code functions, classified for code-search intent and query-code relevance, with annotations from 13 proficient developers. Query4Code leverages Large Language Models (LLMs) to annotate datasets, primarily relying on the semantic information embedded within the code~\cite{li2024optimizing}. However, it lacks test cases or validation mechanisms to ensure the accuracy and reliability of the annotations.

Based on these datasets, COIR~\cite{Li2024CoIRAC} emerges as a robust and comprehensive benchmark specifically designed for code retrieval. COIR comprises ten curated datasets spanning eight retrieval tasks across seven diverse domains, offering a versatile evaluation platform for state-of-the-art retrieval models. Its user-friendly Python framework, compatible with popular benchmarks like MTEB~\cite{Muennighoff2022MTEBMT} and BEIR~\cite{Thakur2021BEIRAH}, facilitates seamless integration into research workflows.

Despite the advancements of these code search datasets, several limitations persist. Doc-code datasets oversimplify the search task by focusing on direct comment-to-code mappings, lacking the complexity of real-world queries. Question-code datasets are not designed for code search tasks, emphasizing question-answer context over direct code retrieval. Query-code datasets, though highly relevant, are constrained by their difficulty in constructing and annotating at scale.

\subsection{Code Search Methods}
The existing methods for code search can mainly be divided into two mainstreams: information retrieval (IR) based methods and deep learning (DL) based models. IR-based methods extract keywords from queries and search for these keywords in the code base to retrieve the most relevant code snippets~\cite{bajracharya2006sourcerer, di2023code, zhang2010understanding}. However, standard information retrieval methods have drawbacks in the code search domain, as they treat codes as text while programming languages and natural languages differ greatly. Fortunately, DL-based methods receive great attention in NLP and achieve significant breakthroughs in code search research. The CodeSearchNet~\cite{husain2019codesearchnet} dataset paves the way for CodeBERT~\cite{feng2020codebert} and GraphCodeBERT~\cite{guo2020graphcodebert}, both of which are bimodal pre-trained BERT~\cite{devlin2018bert} based models that leverage Masked Language Modeling (MLM) and Replaced Token Detection (RTD)~\cite{clark2020electra} objective. UniXcoder~\cite{guo2022unixcoder} unifies three pre-training designs into one architecture and utilizes AST structure and code comment to enhance the cross-modal alignment. Recent T5~\cite{raffel2020exploring} based models like CodeT5~\cite{wang2021codet5} and CodeT5+~\cite{wang2023codet5+} show outstanding performance in code search, utilizing the text-to-text transformer architecture and pre-training on large code datasets. While the two models achieve high performances in the CSN benchmark, GraphCodeBERT~\cite{guo2020graphcodebert} demonstrates superior results by harnessing data flow encoded in the Abstract Syntax Tree (AST) of code to enrich the structural information during pre-training.
Recent advancements in text embedding models have significantly enhanced the performance of code search methods, particularly in multilingual and long-context retrieval tasks. Among these, jina-embeddings-v3~\cite{sturua2024jinaembeddingsv3multilingualembeddingstask} stands out as a state-of-the-art model with 570 million parameters, achieving superior performance on the MTEB benchmark. Its innovative use of task-specific Low-Rank Adaptation (LoRA)~\cite{Hu2021LoRALA} adapters enables high-quality embeddings for various tasks, including query-document retrieval and text matching, while supporting context lengths of up to 8192 tokens. This model outperforms both proprietary embeddings like OpenAI and Cohere. multilingual-e5-large~\cite{wang2024multilingual,multilingual-e5-large}, which is trained on 1 billion multilingual text pairs and fine-tuned on labeled datasets, is efficient, especially in reducing embedding dimensions without compromising quality, thanks to Matryoshka Representation Learning~\cite{Kusupati2022MatryoshkaRL}. But it does not match the flexibility and performance of jina-embeddings-v3. On the other hand, all-MiniLM-L12-v2~\cite{all-minilm-l12-v2} and all-mpnet-base-v2~\cite{all-mpnet-base-v2} represent efficient, compressed models derived from larger pre-trained Transformers. The all-MiniLM-L12-v2 model, based on MiniLM~\cite{wang2020minilm}, employs deep self-attention distillation to retain high accuracy with reduced parameters, making it suitable for latency-sensitive applications. Similarly, all-mpnet-base-v2, built on MPNet~\cite{song2020mpnet}, combines the strengths of BERT and XLNet~\cite{Yang2019XLNetGA} by leveraging masked and permuted language modeling, achieving competitive results on downstream tasks~\cite{wei2023magicoder}. While these models are effective for general-purpose embeddings, they lack the multilingual and long-context capabilities of jina-embeddings-v3 and multilingual-e5-large, limiting their applicability in complex code search scenarios.
\subsection{Test-Driven Agents}
%

Traditional test generation approaches rely heavily on human expertise, but recent work demonstrates the potential of automated test agents. LLM-based agents have shown promising results in various testing scenarios. For instance, in the domain of deep learning libraries, agents capable of generating diverse and meaningful test inputs are proposed~\cite{deng2024large, baudry2024generative}, alongside other studies exploring LLM-based agents for specialized tasks. For test oracle generation, few-shot prompting is used to generate context-aware assertions for test oracle generation~\cite{nashid2023retrieval, deljouyi2024leveraging}, achieving a high accuracy. In program repair, some studies~\cite{xia2023keep, jiang2023impact, xia2023automated, eladawy2024automated, hidvegi2024cigar} demonstrate the effectiveness of integrating LLMs with iterative validation workflows~\cite{xu2025mantra}. However, existing test-driven agents primarily focus on validating specific code functionality without addressing the semantic challenges in code search systems, where our test-driven annotation approach bridges this gap through systematic validation of query-code relevance~\cite{straubinger2025mutation}.

\begin{figure*}[t]
    \centering
    \includegraphics[width=1.0\textwidth]{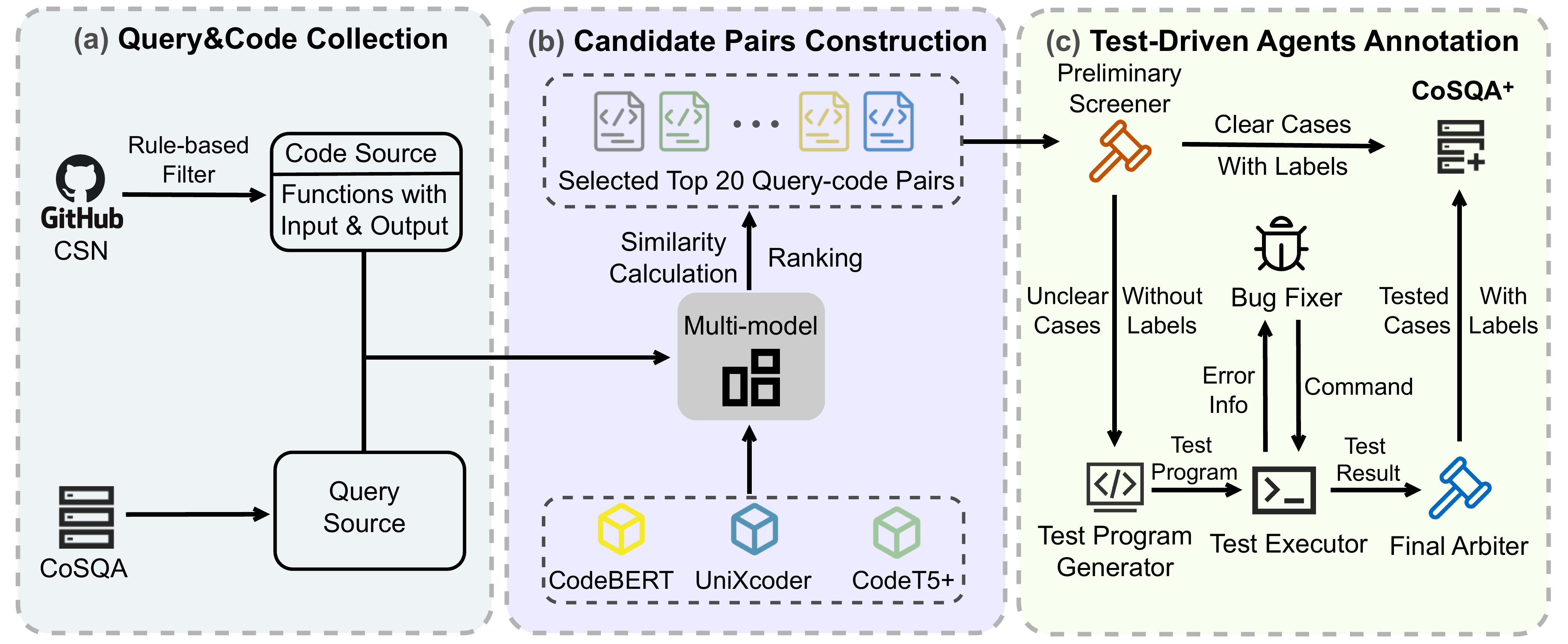}
    \figmargin
    \caption{The construction process of \dataset. (a) Queries sourced from CoSQA and code snippets filtered from CodeSearchNet. (b) Candidate pairs selected based on query-code cosine similarity, computed using a multi-model approach. (c) Pipeline for annotation by test-driven agents. }
    \label{Benchmark Construction}
\end{figure*}

\section{\dataset}

Fig.~\ref{Benchmark Construction} illustrates the benchmark construction process for \textbf{\dataset}. The process starts with query and code collection, in which we leverage the queries from CoSQA and collect python code snippets filtered from CodeSearchNet. In candidate pairs construction, collected code snippets and queries are embedded as vectors to calculate similarity. For each query, the top 20 query-code pairs with the highest similarity are selected as candidates for potential matches. In model annotation, each query-code pair is annotated by test-driven agents with ``1'' or ``0'', indicating whether the code snippet exactly matches the query. These labeled pairs are then added to \dataset.

\subsection{Query and Code Collection}

This section details the reasons why we select CoSQA as the query source and the CodeSearchNet Corpus~\cite{husain2019codesearchnet} serves as the code sources.

\paragraph{Query Source} 

Collected from Microsoft’s Bing search engine logs, CoSQA queries are based on real user queries for python codes~\cite{huang2021cosqa}. We use the queries as the source for the following reasons: 
(1) These queries are drawn directly from real-world programming questions, making them representative of actual developer needs and search patterns. For the practical usability of neural code search, such authentic queries are crucial to creating high-quality datasets~\cite{sun2022importance}. 
\revised{(2) Through heuristic filtering and manual annotation, the queries maintain authentic code search intent while reducing non-code-search-related noise.} This careful curation helps maintain high data quality and reduces noise in the training process~\cite{li2020learning}. By training on these curated real-world queries, code search models can better align with how developers naturally express their programming needs. 


\paragraph{Code Source}

To the best of our knowledge, CodeSearchNet is the only \revised{large-scale, real-world Python function dataset explicitly designed for code search. While other benchmarks (e.g., CoSQA~\cite{huang2021cosqa}, FunSearch~\cite{romera2024mathematical}, AdvTest~\cite{lu2021codexglue}) include Python code, their codebases are derived from CSN — either by reusing its functions directly or by further processing its splits. This makes CSN the de facto origin for Python code-search evaluation. The CodeSearchNet Corpus contains about 6.4 million functions from open-source code spanning six programming languages~\cite{husain2019codesearchnet}.} 

According to our survey, \revised{85.5\% of developers report that functions with explicitly defined inputs and outputs best meet their needs, reflecting real-world developer preferences. Such functions are consistently rated higher in readability, testability, and maintainability~\cite{martin2009clean}. While functions with outputs but no inputs may still be relevant and testable in some contexts, they are often considered poorly designed. These functions typically exhibit unclear logic, rely on implicit global states or hard-coded constants, and are difficult to reuse across different scenarios~\cite{martin2009clean, mcconnell2004code}.}

\revised{To assess data quality and relevance, we carefully reviewed 1,995 functions and found that only 16.25\% of functions without input or output parameters are relevant to typical code search tasks, compared to 80.97\% of functions with both input and output parameters. Including parameter-less functions would therefore introduce noise and degrade overall dataset quality.}

\revised{Therefore, to construct a high-quality Python code corpus from CodeSearchNet’s 1,156,085 functions, we employ a rule-based AST filter to retain only those functions having at least one input parameter and a valid return statement, yielding 653,994 candidates. By concentrating on these clearly specified and testable functions, our dataset preserves sufficient scale, higher quality, and provides a solid foundation for test-driven agents to design and execute meaningful test programs, ultimately improving downstream model quality and supporting accurate code search evaluation.}


\subsection{Candidate Pairs Construction}




Labeling all query-code pairs is impractical. To improve efficiency, we only retain high-confidence pairs before annotation. This section details the construction process of high-confidence query-code pairs.
\paragraph{Construction Steps}
The process of constructing candidate codes for queries in \dataset involves 3 steps aimed at efficiently filtering potential matched codes from a large codebase. This forms the foundation for annotating query-code pairs in the \dataset dataset. 
Firstly, we employ three well-established code embedding models to generate vector representations for both queries and codes.
Then, these embeddings are used to calculate cosine similarity scores, averaging similarity results across the three models. \revised{Cosine similarity is calculated as: 
\[ {cosine\_similarity} = \frac{\mathbf{v}_{\text{query}} \cdot \mathbf{v}_{\text{code}}}{\|\mathbf{v}_{\text{query}}\| \|\mathbf{v}_{\text{code}}\|} \]
where \(\mathbf{v}_{\text{query}}\) and \(\mathbf{v}_{\text{code}}\) are the embedding vectors of the query and code, respectively.} 

\begin{figure}
    \centering
    \includegraphics[width=1.0\linewidth]{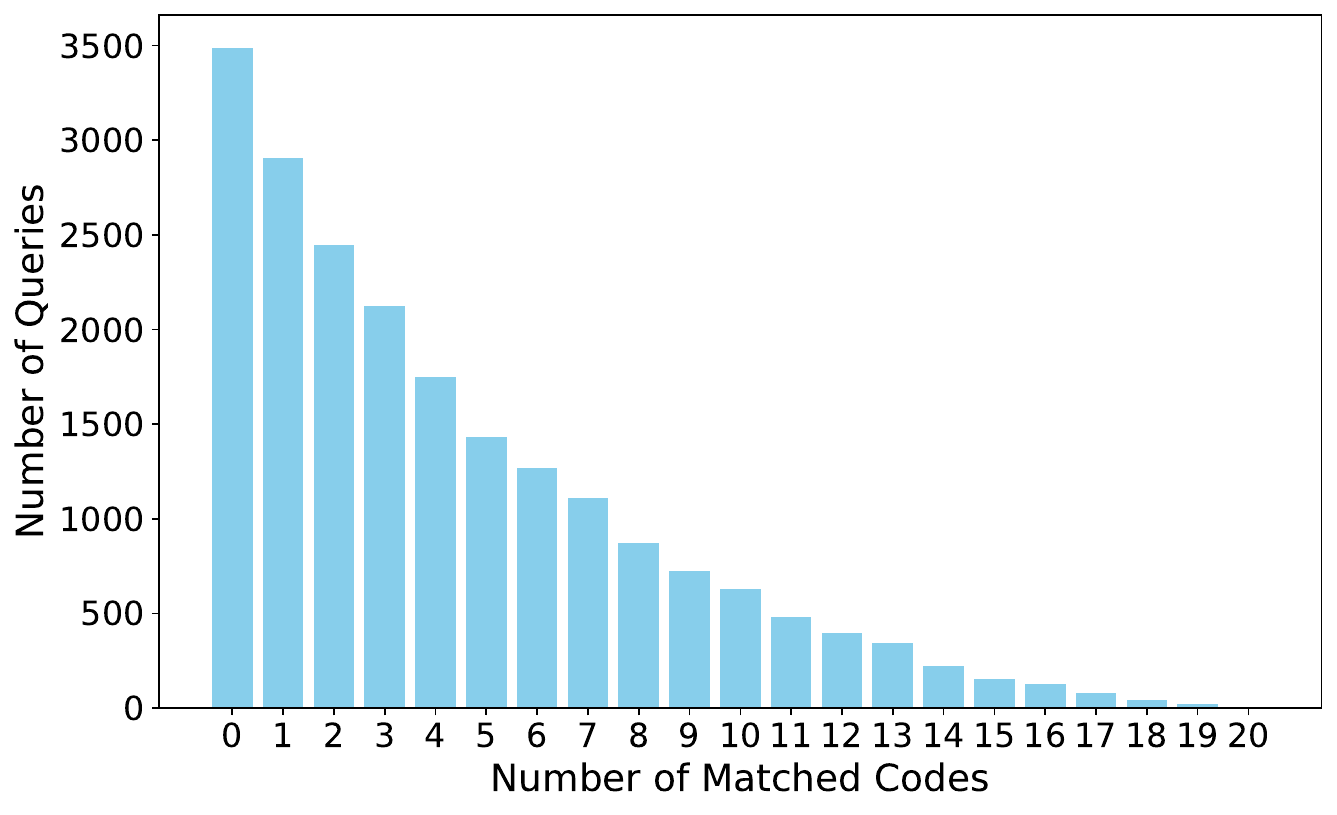}
    \caption{Query counts grouped by number of matched codes}
    \label{fig:query_distribution}
\end{figure}

Finally, the top 20 codes with the highest similarity scores are selected as candidates for each query. Fig. ~\ref{fig:query_distribution} shows a detailed breakdown of query distributions by the number of exactly matched codes.

\paragraph{Multi-model Embedding}

High similarity between query and code is fundamental for a high-quality dataset~\cite{sun2022importance}. During our preliminary assessments, we observe disparities in the similarity scores calculated by different models. Relying solely on one model for computing similarity could introduce bias and skew the dataset unfavorably. This potentially leads to unfair test results of the benchmark. To mitigate this bias, we employ a multi-model approach for calculating the similarity of query-code pairs. We empirically validate this practice. Specifically, we use a single model to select 20 candidate codes for each query and use an ensemble of models, CodeBERT~\cite{feng2020codebert}, UniXcoder~\cite{guo2022unixcoder} and CodeT5+ 110M embedding~\cite{wang2023codet5+}, to select another 20 candidates. Each model embeds each code and query as a vector based on different underlying mechanisms and training datasets. We calculate the overlap between the candidates chosen by the single model and those chosen by the ensemble and analyze the overlap between them. Although bias cannot be completely eliminated, the experiment result in Table~\ref{multi-model-overlap} shows that comprehensively utilizing multiple models for matching can effectively reduce the bias of query-code pairs toward any single model. As can be seen, the proportion of overlapping codes is low, indicating that the codes matched by multiple models for a query do not show a significant preference for any particular single model.

\begin{table}[t]
    \caption{Overlap ratio of candidate codes between individual models and the ensemble of models.}
    \centering
    \begin{tabular}{@{}lr@{}}  \toprule
    Model & Overlap Ratio \\
    \midrule
    CodeBERT & 0.135 \\
    UniXcoder & 0.164 \\
    CodeT5+ embedding & 0.088 \\
    \bottomrule
    \end{tabular}
    \label{multi-model-overlap}
\end{table}

The relatedness between the query and code is quantified by the cosine similarity.

By taking the average similarity score from these three models, we achieve a more balanced and fair assessment of query-code pair quality. 

\begin{figure}[t]
    \centering
    \includegraphics[width=1\linewidth]{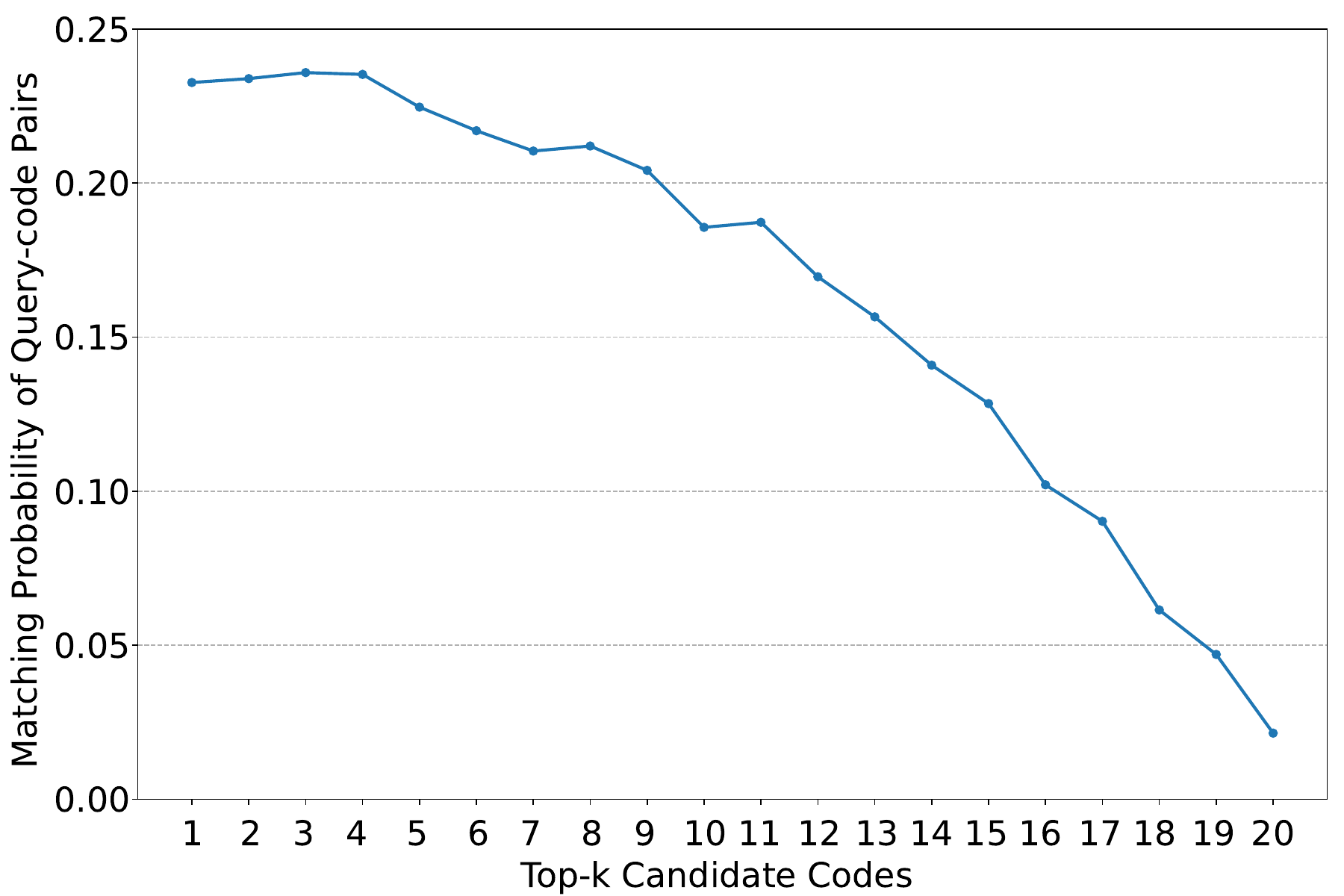}
    \figmargin
    \caption{Probability of identifying matched codes as the number of candidate codes increases}
    \label{TopK}
\end{figure}

\paragraph{Top-20 Code Candidates Selection}
Before constructing the full set of candidate pairs, we randomly select 2,000 queries and annotated them using test-driven agents to determine the optimal number of candidates per query.  The experimental results indicate that the probability of identifying relevant codes decreases significantly as the number of candidates increases (Fig.~\ref{TopK}). As shown in our analysis, when the number of candidates reaches 20, the marginal benefit of further expansion diminishes. This decline can be attributed to limitations of codebase, nature of queries and ability of embedding models. Furthermore, adding just one more candidate code per query results in an additional 20,604 pairs, which would incur substantial costs in terms of both time and financial resources for annotation. Thus, limiting candidates to the top-20 strikes a balance between maximizing the retrieval of relevant codes and controlling annotation costs.

\subsection{Test-Driven Agents Annotation}

To efficiently and accurately annotate the large number of query-code pairs, we design the test-driven annotation agents based on DeepSeek-V3~\cite{deepseekai2024deepseekv3technicalreport}, using the algorithm shown in Algorithm 1. Initially, a preliminary screener determines whether the query clearly matches or does not match the code, or if it is unclear. For clear cases, we directly assign labels based on the initial assessment. For uncertain cases, we determine the labels based on the test execution results and final judgment. In both paths, labels are assigned to indicate whether the code matches the query's requirements (1) or not (0). All labeled query-code pairs are collected to form our dataset \dataset, which consists of pairs that have been rigorously verified either through direct assessment or test-based validation. 

\begin{algorithm}[t]
\caption{Test-Driven Agent Annotation}
\begin{flushleft}
\textbf{Input:} query $q$, code snippet $c$, max\_try\_time \\
\textbf{Output:} label $l \in \{0,1\}$
\end{flushleft}
\begin{algorithmic}[1]
\STATE $assessment \gets \text{preliminary\_screener}(q,c)$
\IF{$assessment = \text{``clearly match''}$}
    \RETURN 1
\ELSIF{$assessment = \text{``clearly non-match''}$}
    \RETURN 0
\ELSIF{$assessment = \text{``unclear''}$}
    \STATE $test\_prog \gets \text{generate\_test\_program}(q,c)$
    \STATE $exec\_res \gets \text{execute\_test}(test\_prog)$
    \STATE $try\_time \gets 0$
    \WHILE{$exec\_res.\text{error} \ \AND \ try\_time \leq max\_try\_time$}
        \STATE $terminal\_commands \gets \text{fix\_test\_program}(test\_prog, exec\_res.\text{error})$
        \STATE $\text{run\_terminal}(terminal\_commands)$
        \STATE $exec\_res \gets \text{execute\_test}(test\_prog)$
        \STATE $try\_time \gets try\_time + 1$
    \ENDWHILE
    \RETURN $\text{final\_arbiter}(q, c, test\_prog, exec\_res)$
\ENDIF
\label{pseudocode_test-agent-modified}
\vspace{-1pt}
\end{algorithmic}
\end{algorithm}

\begin{figure}[t]
    \centering
    \includegraphics[width=0.45\textwidth]{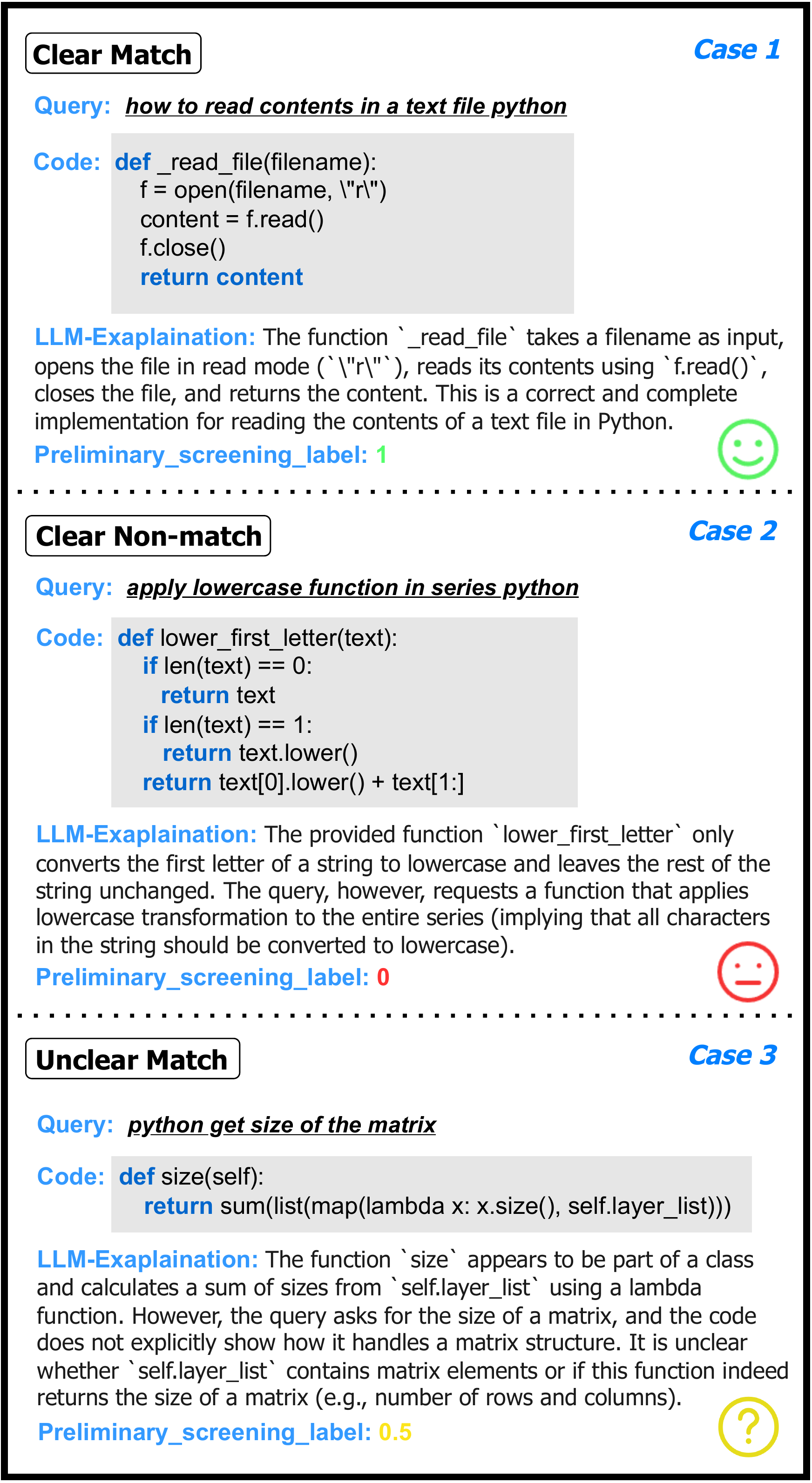}
    \caption{Example illustration of (1) clear match, (2) clear non-match, and (3) unclear cases.}
    \label{fig:clear_example}
\end{figure}

\paragraph{Preliminary Screener}
The evaluation process begins with a preliminary screener that categorizes query-code pairs into three distinct labels: (1) clear match, (2) clear non-match, and (3) unclear, as illustrated in Fig.~\ref{fig:clear_example}. Guided by the prompt instructions, the model assesses whether the code fully addresses the query’s functional requirements. For unambiguous cases, it provides a justification and immediate classification. For ambiguous cases (representing 33\% of candidates), it flags the pair for deeper verification by downstream agents. This triage mechanism optimizes the annotation workflow by resolving straightforward cases upfront while reserving computational resources for ambiguous scenarios.

\paragraph{Test Program Generator}

For ambiguous cases, we employ a test-driven verification methodology. The test program generator constructs executable test suites designed to rigorously validate code functionality. Each test suite includes dependency scaffolding — importing required libraries, defining auxiliary classes, and declaring helper functions — to ensure runtime compatibility. The tests incorporate assert-based validation mechanisms targeting edge cases, functional requirements, and behavioral constraints specified in the query. Considering that a single query may have multiple valid explanations, we generate the test program based on one of the plausible interpretations of the query. This approach transforms ambiguous cases into executable specifications for objective validation. \revised{To be clear, the mere fact that code runs does not guarantee it is relevant to the query. The agents, therefore, generate test programs that execute the code and compare its actual outputs or side effects with the results expected from the query’s intent. Execution logs and assertion outcomes are then passed to the final arbiter, which makes a holistic judgment rather than relying solely on execution success or failure. In this way, verification shifts from a literal understanding of the snippet to functional validation of its behaviour.}

\paragraph{Test Executor and Bug Fixer}

The test executor operates within an isolated Docker environment to ensure safe execution while preventing host system interference. When errors such as \texttt{ModuleNotFoundError} occur, the pipeline initiates a two-stage repair process. First, it attempts direct installation of missing dependencies using standard package managers. If unresolved, the system invokes an LLM to generate context-aware terminal commands for dependency resolution, which are executed iteratively until the error is mitigated or deemed irreparable. After dependency fixes, the test suite is re-executed to validate functionality. Execution traces, assertion outcomes, and error logs are aggregated into a structured report, which is forwarded to the final arbiter for assessment. This approach enhances the reliable verification of ambiguous cases while maintaining environmental integrity, achieving an executable program ratio of 83.67\%.

\begin{figure}
    \centering
    \includegraphics[width=0.85\linewidth]{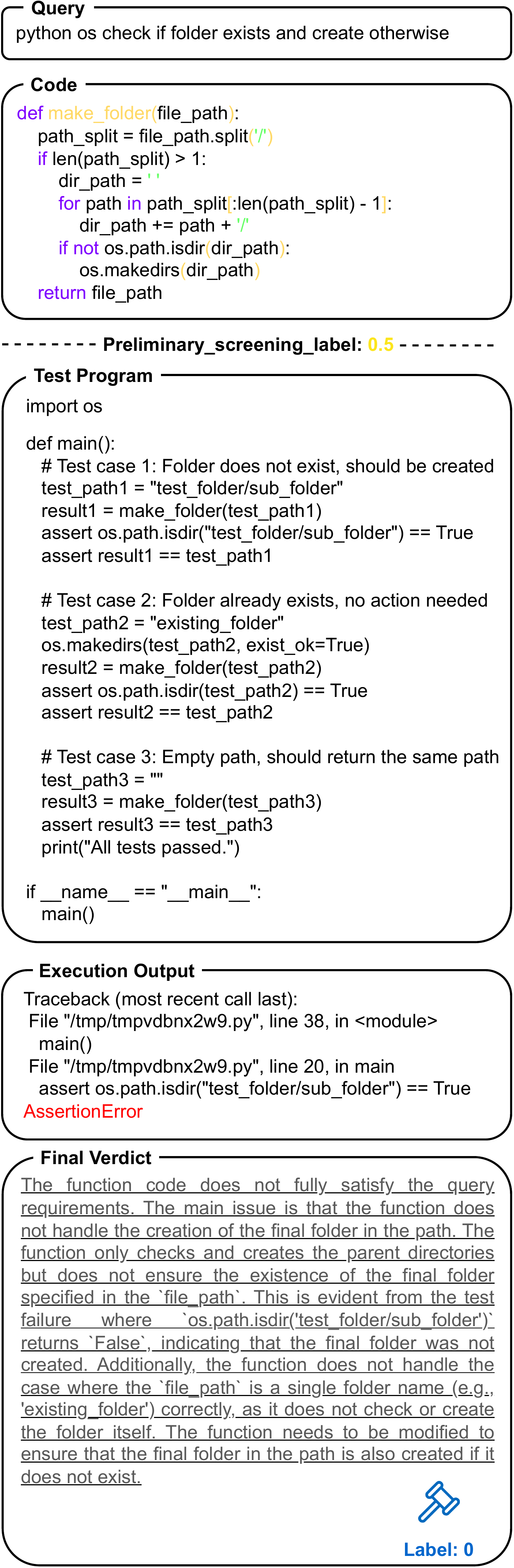}
    \caption{An example for handling an unclear case through test program generation, test execution, and final arbitration.}
    \label{fig:final_arbiter}
\end{figure}

\paragraph{Final Arbiter}
The framework ultimately concludes with a final arbiter stage, where the agent employs a specialized prompt to assess if the generated code fully satisfies the query. This assessment is based on both the test program produced and the outcomes of its execution. This methodology shifts verification from theoretical code inspection to empirical observation, empowering the agent to establish evidence-based conclusions by systematically evaluating actual runtime behavior rather than relying exclusively on static structural analysis and semantic analysis. Fig.~\ref{fig:final_arbiter} illustrates the process of handling an unclear case. 

\subsection{Benchmark Statistics}
In total, 412,080 labeled query-code pairs are retained. Table~\ref{statistic} shows the statistics of our \dataset benchmark.


\begin{table}
    \centering
    \caption{Statistics of \dataset. "\#" denotes the count, "Code avg. len." is the average number of characters per code snippet, and "Exec. rate" represents the percentage of test programs that executed successfully.}
    \begin{tabular}{ccccc} \toprule 
           \# of pairs&  \# of code&  Code avg. len.& Avg. test cases & Exec. rate \\ \midrule 
           412,080&   132,952&  378.052 & 4.31            & 83.67\%\\ \bottomrule
    \end{tabular}
    \tabmargin
    \label{statistic}
\end{table}


\subsection{Cost Analysis}
For data annotation, a total of 412,080 data samples are labeled by Deepseek-V3 at a cost of approximately \$257, resulting in an average cost of \$0.00062 per sample. The labeling time varies significantly based on complexity: clear cases require approximately 10 seconds per sample, while non-clear cases may take up to 50 seconds, depending on API rate limits, dependency installations, and network latency. However, the annotation process benefits from full concurrency across multiple queries, enabling rapid completion when computational resources are sufficient. 

In contrast, manual annotation by Python experts involves substantial time investment. On average, a Python expert spends 8 minutes per sample due to extensive research and analysis. For test-driven annotation, where experts design tailored test programs for code-query pairs, the time increases to 34 minutes per sample, with complex functions exceeding one hour. 

\subsection{Informed consent}

This study involved human participants. Informed consent was obtained from all participants prior to their involvement in the study. Participants were informed about the purpose of the research, the procedures involved, the potential risks and benefits, and their right to withdraw from the study at any time without penalty. All data collected were anonymized to ensure participant confidentiality.

\section{Evaluation}

In this section, we conduct experiments with corresponding analysis to answer the following research questions (RQs):


\textbf{RQ1}: Is the quality of \dataset higher than that of CoSQA?

\textbf{RQ2}: Can our test-driven agents surpass Python experts and LLMs in the annotation task?

\textbf{RQ3}: How effective are the existing approaches in multi-choice code search on our benchmark?

\textbf{RQ4}: How does each agent contribute to the performance of test-driven agents?

\revised{\textbf{RQ5}: Can the test-driven agent generalize across different programming languages?}





\subsection{RQ1: Quality Comparison between \dataset and CoSQA}

We design two experiments to verify whether \dataset offers superior code quality and training performance compared to CoSQA \revised{by examining how well models can learn from these datasets. Here, we treat CoSQA and \dataset as training datasets rather than benchmark suites, focusing on their effectiveness for model training rather than their utility as evaluation benchmarks. Note that this dual functionality — serving as both training corpus and evaluation benchmark — follows established precedent in code search research, where datasets like CodeSearchNet~\cite{husain2019codesearchnet}, CoSQA~\cite{huang2021cosqa} and StaQC~\cite{yao2018staqc} similarly function in both capacities. For this specific comparison, we focus on the training aspect to assess dataset quality and learning effectiveness.}

\paragraph{Metrics}
Following previous work, we employ two metrics to evaluate the performance of code search models: \revised{MAP and MRR~\cite{feng2020codebert}. MAP measures the overall ranking quality by averaging precision scores across all relevant results for each query. For a given query \(q_i\) with \(R_i\) relevant results, we calculate Average Precision (AP) as:}

\revised{\begin{equation*}
AP@k = \frac{1}{R_i}\sum_{j=1}^{k} P(j) \cdot rel(j)
\end{equation*}}

\revised{where \(P(j)\) is the precision at position \(j\), and \(rel(j)\) is an indicator function that equals 1 if the result at position \(j\) is relevant. The final MAP score is obtained by averaging AP across all queries:}

\revised{\begin{equation*}
MAP = \frac{1}{N}\sum_{i=1}^{N} AP@k_i
\end{equation*}}

\revised{where \(N\) is the total number of queries. This metric is particularly suitable for code search as it properly handles cases where multiple correct results may exist for a query.}

MRR evaluates the average reciprocal of the rank at which the first relevant code appears:
\begin{equation*}
MRR = \frac{1}{|N|} \sum_{i=1}^{|N|} \frac{1}{rank_{i}}
\end{equation*}
where |N| is the number of queries and rank\_i is the rank of the first relevant code for the i-th query.

\paragraph{Experimental Setup}
In the first experiment, we randomly select 1000 query-code pairs labeled "1" from \dataset and CoSQA, ensuring the queries are identical. For \dataset, a query may have multiple correct codes, from which we randomly select one that differs from the corresponding query's code in CoSQA. These queries are then presented to experienced programmer volunteers, who are asked to choose the code that satisfies more demand of the query. Finally, we calculate the proportion of better codes from \dataset and CoSQA. 

For the second experiment, We fine-tune CodeBERT, UniXcoder and CodeT5+ embedding with \dataset and CoSQA separately, testing the models on the CSN Python with \revised{MAP} and MRR as evaluation metrics. From the CSN benchmark we use for validation, known as CSN99, we extract Python test samples and build 399 query-code pairs for 95 queries, referred to as CSN Python. The queries in CSN99 are realistic queries sourced from Bing, carefully selected and verified by humans. In our training process, the model is trained using individual query-code pairs, similar to traditional one-to-one code search models. 
The \dataset dataset is divided into training, validation, and test sets in an 8:1:1 ratio, while CoSQA is split according to its original division. 




\begin{table}[t]
\caption{Performance of code search models fine-tuned with different data. The test set is CSN Python.}
\centering
\setlength{\tabcolsep}{2.9pt}
\begin{tabular}{ccccc}
\toprule
 \multirow{2}{*}{Model} & \multirow{2}{*}{Metric} & \multicolumn{3}{c}{Fine-tune Dataset} \\
 \cmidrule(lr){3-5}
& & None & CoSQA & \dataset \\
\midrule
\multirow{2}{*}{CodeBERT} & MRR & 0.952 & 0.939 & \textbf{0.966} \\
& MAP@10& 0.935& 0.900& \textbf{0.938}\\
\multirow{2}{*}{UniXcoder} & MRR & 0.891 & 0.951 & \textbf{0.954} \\
& MAP@10& 0.828& 0.912& \textbf{0.926}\\
CodeT5+ & MRR & 0.794 & 0.938 & \textbf{0.940} \\
embedding & MAP@10& 0.737& 0.910& \textbf{0.928}\\
\bottomrule
\end{tabular}
\label{finetune}
\end{table}

\paragraph{Results and Analysis}
After the volunteers compare the two datasets, we come to the result that out of the 1000 selected queries, 60.3\% of the better codes are from \dataset, while 39.7\% are from CoSQA. This indicates that our \dataset is superior in terms of code quality and the degree of matching between queries and codes. 
As shown in Table~\ref{finetune}, the three models fine-tuned with \dataset outperform those fine-tuned with CoSQA, which indicates better training performance of \dataset. Notably, the improvement in MAP@10 is more significant than the improvement in MRR, highlighting the effectiveness of \dataset in training the model to retrieve multiple relevant code matches for a single query.

\begin{center}
    \begin{myboxc} \textbf{RQ1 Summary: } 
    \dataset has better quality than CoSQA. Fine-tuning CodeBERT, UniXcoder and CodeT5+ embedding with \dataset yields better performance than with CoSQA, especially for multiple relevant codes.
    \end{myboxc} 
\end{center}

\subsection{RQ2: Performance of Agent-based Annotator}
We design this experiment to assess whether our agent-based annotation approach can achieve comparable or superior quality to traditional human annotation and single LLM annotation.

\paragraph{Experimental Setup}


To establish a reliable ground truth, \revised{we recruit three experienced Python developers with 3-5 years of professional programming experience and computer science backgrounds to serve as expert annotators. These experts first independently design test programs for 1000 randomly selected query-code pairs from \dataset, following a standardized evaluation protocol that required: (1) Functional correctness verification through executable test cases. (2) Edge case coverage assessment. (3) Query requirement satisfaction analysis. They then annotate the selected pairs based on test execution results.}

The final ground truth labels are determined by majority voting among these three experts, \revised{with an initial inter-annotator agreement measured by Krippendorff's alpha coefficient, which is particularly suitable for assessing agreement among multiple annotators~\cite{krippendorff1980krippendorff}. In cases of disagreement, experts engage in discussion sessions to reach a consensus.}
We then evaluate three different annotation approaches against this established ground truth: 
(1) human annotators without test program verification;
(2) our agent-based annotator;
(3) single LLM annotators (including Deepseek-V3, O1-mini, and Claude-3.5-Sonnet), \revised{each provided with identical query-code pairs and evaluation criteria.}

\paragraph{Results and Analysis}

\begin{table}[t]
    \caption{Accuracy comparison of different annotation methods against ground truth.}
    \centering
    \begin{tabular}{@{}lr@{}}  \toprule
    Method & Accuracy \\
    \midrule
    Agent-based Annotator & 0.939±0.020 \\
    Human Wo. Test Program & 0.891±0.026 \\
    Deepseek-V3 & 0.883±0.027 \\
    O1-mini & 0.881±0.027 \\
    Claude-3.5-Sonnet & 0.863±0.029 \\
    \bottomrule
    \end{tabular}
    \label{annotation-performance}
\end{table}

As shown in Table~\ref{annotation-performance}, our agent-based annotator achieves \revised{93.9\%} accuracy compared to the ground truth, significantly outperforming human annotators without test program verification (89.1\%), Deepseek-V3 (88.3\%), O1-mini (88.1\%) and Claude-3.5-Sonnet (86.3\%). This high accuracy demonstrates that our automated agent-based approach can effectively match the quality of expert annotations with test program verification. \revised{To ensure appropriate interval estimation for our data characteristics, we adopt the Clopper–Pearson exact method with 95\% confidence intervals for accuracy metrics in TABLE~\ref{annotation-performance}. Clopper–Pearson exact method is well-suited for moderate sample sizes and high sample proportions (i.e., proportions close to 0 or 1).}

Importantly, about 83\% of the test programs generated by our agent are successfully executable, indicating the reliability of our test program generation approach. The high executability rate indicates that most query-code pairs can be verified through actual program testing rather than relying solely on semantic analysis. 

The superior performance of our agent-based annotators can be attributed to two key factors: 1) The test program provides concrete evidence of whether the code satisfies the query requirements, reducing ambiguity in the annotation process; 2) The systematic verification through test execution helps eliminate human bias and inconsistency that often occur in manual annotation.

In terms of inter-annotator agreement measured by Krippendorff's alpha coefficient, the experts with test programs show strong consistency with a coefficient of \revised{0.82}, notably higher than human annotations without test program verification (\revised{0.60}), which suggests that test programs help reduce subjectivity in the annotation process.

\begin{center}
    \begin{myboxc} \textbf{RQ2 Summary: } 
    Our agent-based annotator achieves \revised{93.9\%} accuracy, outperforming both traditional human annotation and single LLM approaches. The high executability rate (83.67\%) of generated test programs demonstrates the reliability of our automated annotation approach.
    \end{myboxc}
\end{center}

\subsection{RQ3: Existing Methods Performance on \dataset}
This experiment aims to evaluate the effectiveness of different models in accurately retrieving code snippets that match natural language queries on the \dataset benchmark.

\paragraph{Task}

The natural language code search task is a text retrieval problem. Given a natural language query \(q_i\) and a collection of code snippets \(C = \{c_1, \ldots, c_n\}\), the goal is to identify all code snippets 
\[
\{c^*_j \mid c^*_j \in C \text{ and } c^*_j \text{ exactly matches } q_i\}
\]
that exactly match the query.

\paragraph{Experimental Setup}
The effectiveness of semantic code search models is evaluated on the \dataset benchmark using \revised{MAP@10} as the primary metric. \revised{MAP@10} is particularly suitable for our scenario, where multiple correct code snippets may exist for a query, as it measures how well a model ranks all relevant results rather than focusing only on the first correct match. Additionally, we report Mean Reciprocal Rank (MRR) to assess the ability to rank the first relevant result highly.

\revised{To provide a more comprehensive evaluation, we also include NDCG~\cite{wang2013theoretical, Li2024CoIRAC, wang2020trans3transformerbasedframeworkunifying, rahman2019automatic} (Normalized Discounted Cumulative Gain) and Recall as additional metrics. Although NDCG is typically used for graded relevance judgments, in our case where relevance is binary (either matching or not), it still serves as a useful metric by emphasizing the importance of ranking relevant results higher in the list.} Meanwhile, Recall measures the model's ability to retrieve all relevant code snippets from the dataset, ensuring comprehensive coverage of correct results. Together, these metrics offer a robust and multi-faceted assessment of the model's performance in semantic code search tasks. 
The tested methods include both traditional information retrieval techniques and advanced deep learning-based models. Specifically, we compare a traditional Bag of Words (BoW) approach~\cite{zhang2010understanding} with state-of-the-art semantic code search models, namely: CodeBERT~\cite{feng2020codebert}, UniXcoder~\cite{guo2022unixcoder}, and CodeT5+ embedding~\cite{wang2023codet5+}. In addition, we evaluate several recent sentence embedding models, including Jina-embedding-v3~\cite{sturua2024jinaembeddingsv3multilingualembeddingstask}, All-MiniLM-L12-v2~\cite{all-minilm-l12-v2}, All-mpnet-base-v2~\cite{all-mpnet-base-v2}, and Multilingual-e5-large~\cite{multilingual-e5-large}.

\paragraph{Results and Analysis}

\begin{table}[t]
    \caption{Performance of Various Methods on \dataset Using NDCG@10, MRR, MAP and Recall Metrics.}
    \centering
    \begin{tabular}{@{}lllll@{}}  \toprule
        Method                 & NDCG@10  & MRR & MAP & Recall     \\ \midrule
        BoW                    & 0.037  & 0.027 & 0.027 & 0.016 \\
        CodeBERT               & \textbf{0.232}  & \textbf{0.175} &\textbf{0.164} &\textbf{0.149}\\
        UniXcoder              & 0.187  & 0.140 & 0.132 & 0.110 \\
        CodeT5+ embedding      & 0.179  & 0.136 &  0.127  & 0.101 \\
        Jina-embedding-v3       & 0.225  & 0.171  & 0.160 & 0.145 \\
        All-MiniLM-L12-v2      & 0.214  & 0.163  & 0.153  & 0.135 \\
        All-mpnet-base-v2      & 0.227  & 0.172  & 0.162  & 0.145 \\
        Multilingual-e5-large  & 0.218  & 0.166  & 0.156  & 0.136 \\
        \bottomrule
    \end{tabular}
    \label{NDCG and MRR on CoSQA+}
\end{table}
While traditional information retrieval methods, such as BoW, are computationally efficient, Table~\ref{NDCG and MRR on CoSQA+} demonstrates that they struggle to capture the semantic subtleties needed to accurately map natural language queries to relevant code snippets, particularly when multiple valid solutions exist. Among the deep learning-based models, CodeBERT achieves the highest performance with NDCG@10 of 0.232, MRR of 0.175, MAP of 0.164 and Recall of 0.149. Other embedding-based models also show substantial improvements over traditional methods, highlighting the benefits of modern deep learning architectures for semantic code search. This strong performance can be attributed to their encoder-only architecture and extensive pre-training on large-scale paired natural and programming language corpora, especially Python, which enhances their code understanding capabilities.
The embedding models originate from different time periods and vary in parameter sizes. Surprisingly, CodeBERT outperforms the latest embedding models, which may be attributed to the fact that CodeBERT tested here is fine-tuned specifically for code search tasks, whereas the latest embedding models are designed for general-purpose tasks. Some smaller-parameter embedding models demonstrate impressive performance. For instance, all-mpnet-base-v2, with only 109M parameters, and all-MiniLM-L12-v2, with a mere 33M parameters, still exhibit remarkable performance, surpassing many models with larger parameter sizes. This represents a significant advancement in embedding models in recent years, providing a solid foundation for the practical implementation of semantic search based on large models.


\revised{Regarding the evaluation metrics (MRR, NDCG, MAP): these inherently assign high weight to the top-ranked results. Consequently, the most critical factor influencing a model's score is its ability to rank correct candidate snippets at the very top positions. The internal relative order of matched code snippets in the search results has no influence on the score. A model is considered better on \dataset if it can find more matched code and position them at the top of the results.}

From the evaluation results, the performance differences among embedding models do not seem significant. However, due to the richness of the \dataset, which effectively eliminates random factors, even subtle differences in metrics can reflect the overall capability gap between models. Moreover, embedding models generally perform poorly on our benchmark. From the perspective of model design, embedding models are mainly used for retrieval tasks. To improve inference speed, their weight parameters are usually small, limiting their ability to deeply understand query sentences and code. 

The \dataset is challenging in two main aspects: First, the queries originate from CoSQA, which are real user queries filtered from the Bing search engine. These queries are often expressed casually and may contain spelling or grammar errors. Second, the codebase used for retrieval is not only large in scale but also contains many confusing code snippets.

\begin{center}
    \begin{myboxc}
      \textbf{RQ3 Summary: } 
      The evaluation on \dataset clearly indicates that deep learning-based models significantly outperform traditional retrieval methods for semantic code search. Notably, CodeBERT leads the performance ranking by achieving the highest scores in both MAP@10 and MRR, while the alternative embedding-based models also offer competitive results.
    \end{myboxc}
\end{center}

\subsection{RQ4: Contributions of Different Components in Test-Driven Agents}
This experiment is to assess the contribution of each component in our test-driven agent framework.
\paragraph{Experimental setup}
The final test-driven agents comprise the following components: Preliminary Screener (PS), Test Program Generator (TPG), Test Executor (TE), Bug Fixer (BF), and Final Arbiter (FA). We divided the agent into different configurations by combining these components in various ways. For each configuration, we compared the predicted labels with the ground truth labels (obtained from human evaluation based on the test programs) to calculate the accuracy. Additionally, we compare the Simulated Executor with both the Test Executor and the Bug Fixer to evaluate whether the large language model can accurately simulate execution without actual interpretation. Unlike the Test Executor, which runs the test program using a Python interpreter, the Simulated Executor leverages the LLM's reasoning capabilities to emulate execution and directly predict the outcome.

\paragraph{Result and Analysis}
Table~\ref{tab:components} shows the accuracy for different configurations.

\begin{table}[ht]
\centering
\begin{tabular}{l r}
\toprule
\textbf{Configuration} & \textbf{Accuracy} \\
\midrule
Single LLM & 0.883 \\
PS + FA & 0.873 \\
PS + TPG + FA & 0.852 \\
PS + TPG + TE + FA & 0.854 \\
PS + TPG + Simulated Executor + FA & 0.853 \\
PS + TPG + TE + BF + FA & 0.939 \\
\bottomrule
\end{tabular}
\caption{Accuracy of Various Test-Driven Agent Configurations (Abbreviated)}
\label{tab:components}
\end{table}

The results indicate that the baseline Single LLM achieves an accuracy of \revised{0.883}. When the system uses only the Preliminary Screener in combination with the Final Arbiter, the accuracy slightly drops to \revised{0.873}. Incorporating the Test Program Generator alongside these yields an accuracy of \revised{0.852}, suggesting that generating test programs without execution may introduce noise or irrelevant information, which can harm performance~\cite{zhang2021understanding}. Adding the Test Executor to the configuration improves the performance marginally (\revised{0.854}), and using a Simulated Test Executor does not provide a significant advantage (\revised{0.853}). This aligns with the observation that simulated execution cannot fully capture the complexity of the real compile-and-run process, potentially introducing inaccuracies or misleading feedback. Notably, the inclusion of the Bug Fixer increases the accuracy substantially to \revised{0.939}. This enhancement demonstrates that the Bug Fixer is crucial for processing the test programs effectively, as it enables the system to generate real execution results that are vital for improving prediction accuracy.

\begin{center}
    \begin{myboxc}
      \textbf{RQ4 Summary: } 
     The Bug Fixer is the most important component in the test-driven agent framework as it makes the test program executable and yields effective test results. However, it relies on the preceding steps, such as generating the test program. Simulated execution results cannot replace the benefits of real execution feedback.
    \end{myboxc}
\end{center}

\revised{\subsection{RQ5: Cross-Language Generalization of the Test-Driven Agent}}

\revised{This experiment is to explore the generalizability of the test-driven agent's annotation performance beyond Python.}

\revised{\paragraph{Experimental Setup}}

\revised{We collect and annotate query-code pairs from CodeSearchNet (CSN) for PHP, Java, and Go to benchmark the test-driven and single-LLM annotation methods. We select these languages because (1) they rank among the top ten most-used languages, ensuring real-world relevance~\cite{itransition2025}, and (2) they represent distinct paradigms: dynamically typed scripting (PHP), statically typed object-oriented (Java), and statically typed procedural/concurrent (Go).
We build the benchmark in two steps. First, we extract unannotated query-code pairs from CSN for each language. Second, we label every pair using the same pipeline described in RQ2 to ensure consistent evaluation.
To evaluate cross-language performance, we adapt the pipeline in Algorithm 1 to each language. We slightly refine the prompts and configure separate execution environments for PHP, Java, and Go.
We then compare two annotation methods against the human-labeled ground truth: (1) the test-driven agent based on DeepSeek-V3 and (2) the single-pass LLM annotator (DeepSeek-V3). Accuracy is the metric, defined as the fraction of labels that match human annotations.}

\revised{\paragraph{Result and Analysis}}

\revised{The results reveal a consistent pattern: the test-driven agent achieves higher annotation accuracy than single LLM annotators across all three non-Python languages, with significant improvements observed in PHP (84.87\% vs. 72.70\%, +12.17\%), Java (79.96\% vs. 69.59\%, +10.37\%), and Go (81.93\% vs. 73.49\%, +8.44\%). This performance improvement indicates that the test-driven framework, by incorporating functional validation and error correction mechanisms, effectively enhances cross-language code understanding capabilities, outperforming the baseline of single LLMs.}

\revised{Our test-driven agent consistently outperforms the single LLM baseline in accuracy. This highlights a key advantage: our approach mitigates the limitations of pre-trained models like DeepSeek-V3. Such models may exhibit inherent limitations stemming from their training data composition, which often over-represents certain languages or paradigms. This can reduce their robustness when handling syntax and features specific to other languages. For instance, PHP’s dynamic typing quirks pose interpretive challenges for the single LLM baseline. However, the test-driven agent’s structured validation process effectively addresses the issues, resulting in clearer improvements. Notably, PHP shows the largest accuracy gain (12.17\%) compared to the baseline, which could stem from its unique blend of scripting and web-focused features—elements that demand stricter contextual validation.}

\begin{center}
    \begin{myboxc} \revised{\textbf{RQ5 Summary: } 
    The test-driven agent outperforms single LLM annotators across PHP, Java, and Go with an average 10.33\% accuracy gain, showing strong cross-language generalization.}
    \end{myboxc} 
\end{center}

\section{Discussion}

\subsection{Test Misjudgment}
Previous studies show that as the size of the codebase increases, the challenges faced by embedding models also increase~\cite{liu2024codexgraphbridginglargelanguage, nie2025textembeddingmeetslarge}. This conclusion aligns with intuition, but with the analysis of other datasets and the construction of \dataset, we find that the decline in performance is not only due to the need to exclude more confusing codes but also because of the increased likelihood of test misjudgments. For example, an LLM may select a piece of code that meets the query requirements, but this query-code pair is not included in the candidate code set. As a result, this correct selection is misjudged as incorrect.

\begin{figure}
    \centering
    \includegraphics[width=0.5\linewidth]{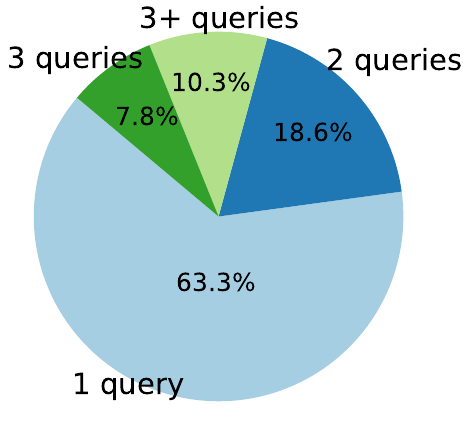}
    \caption{Distribution of codes by the number of queries they satisfy}
    \label{fig:code_counts_by_query}
\end{figure}

When constructing code search datasets, query sets and code sets are typically extracted from original query-code pairs for test. However, the number of all possible query-code pairs (similar to a Cartesian product) formed between the query set and the code set far exceeds the number of original pairs. Existing methods assume that the code of a pair outside the original pairs cannot correctly match the query, but this assumption is not entirely valid. As shown in Fig.~\ref{fig:code_counts_by_query}, in \dataset, excluding codes without satisfied query, 36.7\% of codes can satisfy two or more queries. Constructing a dataset completely free of misjudgments is almost impossible, as it would require checking all possible pairs. As can be seen by combining Fig~\ref{fig:query_distribution} and Fig.~\ref{fig:code_counts_by_query}, the approach adopted in \dataset, where 20 candidate codes are annotated for each query, helps mitigate these testing misjudgments. Another possible measure is to invoke test-driven annotation agents when a code search method selects a previously unseen query-code pair to determine whether the selected code matches the query. However, this approach may introduce significant computational overhead, especially when the candidate code repository is very large.

\subsection{Inherent Ambiguity in Query and Code Labels}
A major challenge in code search is the inherent ambiguity in queries and code. Queries are often vague or poorly phrased, leading to multiple interpretations, while code snippets may be incomplete or abstracted, making correctness subjective. For instance, a query like "how to check if a folder exists and create it if not" can be interpreted differently based on assumptions about paths or edge cases. Similarly, code may solve the problem partially or in a non-standard way, complicating strict labeling ("exact match"). Test-driven annotation helps reduce ambiguity by generating executable test programs aligned with the query's intent. However, variability in query interpretation and incomplete code still introduce labeling challenges. Future improvements will require clearer queries and more complete code snippets to minimize ambiguity.

\subsection{Ethical Consideration}
In this work, we prioritize ethical considerations by the responsible use of data and methodologies. The dataset sources, CoSQA and CodeSearchNet, are publicly available under permissible licenses, mitigating privacy concerns. Our test-driven agents are designed to minimize human bias by automating code evaluation, thus enhancing fairness and consistency. Additionally, we openly release \dataset to promote transparency and reproducibility in code search research. While our approach aims to improve code search efficiency, we acknowledge the potential for misuse in generating or retrieving malicious code. We encourage the community to adopt ethical guidelines when using our benchmark and tools, ensuring that their application aligns with positive social impact.

\revised{
\subsection{Failure Case Analysis}
Our test-driven agents achieve 93.9\% accuracy, but understanding the remaining 6.1\% failure cases is crucial for future improvements. We analyze ten false-positive and false-negative cases and identify four primary root causes: (1) Ambiguous queries (30\% of cases) where natural language descriptions like "python fabric verbose logging" have multiple valid interpretations (e.g., generic logging functionality vs. Fabric library-specific SSH logging), leading to mismatched test generation; (2) Human annotation biases (30\% of cases) where annotator subjectivity leads to incorrect judgments, such as being misled by superficial keyword matching while agents correctly identify functional mismatches;
(3) Semantic misalignment (20\% of cases), where agents fail to capture the precise intent behind technical terms or behavioral descriptors; (4) Validation misalignment (20\% of cases), where test programs validate code functionality rather than query requirements, such as testing all features of a function instead of the specific query need. The analysis reveals that most failure cases stem from inherent limitations in natural language understanding rather than systematic flaws in our test-driven approach. Query ambiguity remains the most significant challenge, suggesting that future work should focus on multi-interpretation test generation or query clarification mechanisms. Importantly, our agents sometimes perform better than ground truth labels, indicating that test-driven verification can identify annotation failure cases in human-labeled datasets.}

\revised{
\subsection{Validation Strategy and Scalability Constraints}}
\revised{Complete human verification of large-scale datasets faces fundamental scalability barriers. Expert annotation with test programs requires 34 minutes per pair, making exhaustive validation of 412,080 pairs economically prohibitive at approximately 234,000 person-hours. Our sampling-based validation strategy addresses this constraint while maintaining methodological rigor. To strengthen validation beyond our 1,000-sample ground truth evaluation, we implemented an additional verification mechanism: a specialized checker agent using DeepSeek-V3 that performs secondary validation of our test-driven annotations. This checker evaluates the complete annotation pipeline output, including preliminary screening results, test programs, execution outcomes, and final judgments. The checker achieves high agreement with human expert annotations (Krippendorff's $\alpha$ = 0.914) and, as ground truth, grants the test-driven agents 92.8\% accuracy.
The high inter-annotator agreement between our checker and human experts demonstrates that automated verification can reliably assess annotation quality at scale. This checker-based validation mechanism enables quality assurance across the entire dataset while maintaining economic feasibility, offering a practical solution to the scalability constraints of traditional manual verification approaches.}


\section{Conclusion}
In this paper, we present \dataset, the first large-scale multi-choice code search dataset with 412,080 pairs. Our dataset expands from single-choice to multi-choice query-code matching pairs, enhancing its suitability for real-world code search tasks. Through rigorous experiments, \dataset demonstrates superior quality and training performance compared to existing benchmarks like CoSQA. The test-driven annotation approach achieves a high accuracy of \revised{93.9\%}, outperforming \revised{not only traditional human annotation and single-LLM methods but also demonstrating strong cross-language generalizability.} Experiments on \dataset show that CodeBERT is better suited for multi-choice code search tasks. We believe our agents-annotated \dataset benchmark will offer insights for other research that involves natural language and code, and for evaluations on a multi-choice code search task.

\newpage
\revised{\section{Code Search Behavior Survey Report}}

\revised{We conducted a survey with 200 valid respondents to understand their code search behaviors.} 

\revised{\subsection{Part I: Basic Information}}

\revised{\textbf{Question 1: What is your highest level of education?}
\begin{itemize}
    \item A. Associate Degree: 2 (1.0\%)
    \item B. Bachelor's Degree: 175 (87.5\%)
    \item C. Master's Degree: 15 (7.5\%)
    \item D. Doctoral Degree: 8 (4\%)
\end{itemize}}

\revised{\textbf{Question 2: How many years of programming work experience do you have?}
\begin{itemize}
    \item A. Less than 1 year: 47 (23.5\%)
    \item B. 1-5 years: 118 (59.0\%)
    \item C. 5-10 years: 32 (16.0\%)
    \item D. More than 10 years: 3 (1.5\%)
\end{itemize}}

\revised{\textbf{Question 3: Do you have a computer science background (having taken at least 4 computer-related courses such as Data Structures \& Algorithms, Compiler Principles, Operating Systems, etc.)?}
\begin{itemize}
    \item A. Yes: 186 (93.0\%)
    \item B. No: 14 (7.0\%)
\end{itemize}}

\revised{\textbf{Question 4: Is Python one of your primarily used programming languages?}
\begin{itemize}
    \item Yes (with experience selection): 128 (64.0\%)
    \item No: 72 (36.0\%)
\end{itemize}}

\revised{\subsection{Part II: Code Search Behavior}}

\revised{\textbf{Question 5: When performing programming-related tasks, what is your average daily frequency of code searches?}
\begin{itemize}
    \item A. More than 10 times: 82 (41.0\%)
    \item B. 6-10 times: 48 (24.0\%)
    \item C. 3-5 times: 41 (20.5\%)
    \item D. 1-2 times: 12 (6.0\%)
    \item E. Rarely used: 17 (8.5\%)
\end{itemize}}

\revised{\textbf{Question 6: In your daily development, how many code examples do you typically need to reference for one query (such as API usage, algorithm implementation, etc.)?}
\begin{itemize}
    \item A. Usually only need 1: 24 (12.0\%)
    \item B. Usually need 2-3: 133 (66.5\%)
    \item C. Usually need 4-5: 32 (16.0\%)
    \item D. Often need more than 5: 11 (5.5\%)
\end{itemize}}

\revised{\textbf{Question 7: In your daily development, how many web pages returned by search engines do you typically need to open for one query?}
\begin{itemize}
    \item A. Usually only need 1: 26 (13.0\%)
    \item B. Usually need 2-3: 107 (53.5\%)
    \item C. Usually need 4-5: 47 (23.5\%)
    \item D. Often need more than 5: 20 (10.0\%)
\end{itemize}}

\revised{\textbf{Question 8: When conducting code searches, what are the most common scenarios where you need to reference multiple code examples? [Multiple Choice]}
\begin{itemize}
    \item A. API usage methods: 141 (70.5\%)
    \item B. Algorithm implementation approaches: 133 (66.5\%)
    \item C. Performance optimization solutions: 109 (54.5\%)
    \item D. Compatibility handling: 87 (43.5\%)
    \item E. Other: 12 (6.0\%)
\end{itemize}}

\revised{\textbf{Question 9: In actual development, what do you think is the approximate percentage of queries where multiple code examples meet your requirements?}}

\revised{Average response: 63.2\% (total score: 12,635/20,000)}

\revised{\textbf{Question 10: In which of the following situations would you tend to look at multiple code examples? [Multiple Choice]}
\begin{itemize}
    \item A. Learning new technologies or frameworks: 156 (78.0\%)
    \item B. Solving complex problems: 147 (73.5\%)
    \item C. Optimizing existing code: 125 (62.5\%)
    \item D. Handling edge cases: 85 (42.5\%)
    \item E. Ensuring code quality: 89 (44.5\%)
    \item F. Other: 8 (4.0\%)
\end{itemize}}

\revised{\textbf{Question 11: When using code search tools (such as Google, Stack Overflow, GitHub search, etc.), have you encountered the following difficulties? [Multiple Choice]}}
\revised{\begin{itemize}
    \item A. Too many similar examples in search results, difficult to quickly find the most suitable one: 133 (66.5\%)
    \item B. Need to open multiple pages to compare different implementation approaches: 135 (67.5\%)
    \item C. Search results not sorted by practicality/popularity: 108 (54.0\%)
    \item D. Cannot simultaneously display usage methods for different scenarios: 95 (47.5\%)
    \item E. Code examples lack usage scenario descriptions: 86 (43.0\%)
    \item F. Other: 3 (1.5\%)
\end{itemize}}

\revised{\textbf{Question 12: When searching for code, which of the following better matches your search habits?}
\begin{itemize}
    \item A. Only need to find one code snippet that meets requirements: 81 (40.5\%)
    \item B. Need to find more than one code snippet that meets requirements: 119 (59.5\%)
\end{itemize}}

\revised{\textbf{Question 13: When search returns multiple code examples, which of the following aspects do you pay most attention to? [Multiple Choice]}
\begin{itemize}
    \item A. Differences between code examples (e.g. comparisons between different implementation methods): 133 (66.5\%)
    \item B. Relevance of each example (e.g. degree of match with the query problem): 133 (66.5\%)
    \item C. Whether the display order of examples meets expectations (e.g. whether the most relevant ones are ranked first): 108 (54.0\%)
    \item D. Whether examples cover common usage scenarios or boundary conditions: 83 (41.5\%)
    \item E. Other: 2 (1.0\%)
\end{itemize}}

\revised{\textbf{Question 14: When evaluating the quality of code search results, which of the following do you think is more important?}
\begin{itemize}
    \item A. Position where the first suitable result appears: 73 (36.5\%)
    \item B. Average position of the first few suitable results: 89 (44.5\%)
    \item C. Diversity of results: 37 (18.5\%)
    \item D. Other: 1 (0.5\%)
\end{itemize}}

\revised{\textbf{Question 15: Out of 10 queries, what is the proportion of using only code search, only code generation, and using both code search and code generation?}
\begin{itemize}
    \item Only code search (32.55\%) 
    \item Only code generation (29.64\%)
    \item Code search + code generation (37.81\%) 
\end{itemize}}

\revised{\textbf{Question 16: When searching for code examples, which function structure do you prefer?}
\begin{itemize}
    \item A. Functions with both input and output: 171 (85.5\%)
    \item B. Functions with input but no output: 13 (6.5\%)
    \item C. Functions with output but no input: 10 (5.0\%)
    \item D. Functions with neither input nor output: 6 (3.0\%)
    \item Other: 0 (0\%)
\end{itemize}}

\revised{\textbf{Question 17: When searching for code examples, which function structure do you generally end up using?}
\begin{itemize}
    \item A. Functions with both input and output: 174 (87.0\%)
    \item B. Functions with input but no output: 17 (8.5\%)
    \item C. Functions with output but no input: 2 (1.0\%)
    \item D. Functions with neither input nor output: 7 (3.5\%)
    \item E. Other: 0 (0\%)
\end{itemize}}

\revised{\section{DETAILED FAILURE CASE ANALYSIS}
This section provides ten false-positive/ false-negative cases with discussion of root causes encountered during test-driven annotation (Table~\ref{tab:query_code_examples}).}

\revised{\subsection{Ambiguous Query Cases}}
\revised{\textbf{Case 1}: Query "python fabric verbose logging" matched with a basic logging function. The test program incorrectly assumed generic logging functionality, while "fabric" specifically refers to the Python Fabric library for SSH operations. When we refined the query to "python use fabric to verbose logging," the test-driven agent correctly labeled it as non-matching (label: 0), demonstrating the impact of query clarity on annotation accuracy.}

\revised{\textbf{Case 2}: Query "slice every 5 items python" has multiple valid interpretations: taking first 5 items, every 5th item, or chunking into groups of 5. The agent generated tests for one interpretation, missing the actual intent, highlighting the need for multi-interpretation testing strategies.}

\revised{\textbf{Case 3}: Query "lowercase + string object + python" is overly abstract without specific functional requirements. The agent failed to interpret the vague query, generating misaligned tests that missed the actual code's selective lowercasing behavior for boolean values.}

\revised{\subsection{Human Annotation Biases}}
\revised{\textbf{Case 4}: For query "how to get next month python datetime," human annotators labeled a time-interval addition function as incorrect (label: 0). However, the test-driven agent correctly identified that the function doesn't calculate next month properly, as it adds fixed seconds rather than adjusting month/year components.}

\revised{\textbf{Case 5}: For query "python string to cstr," human annotators incorrectly labeled a JSON serialization function as matching (label: 1). The test-driven agent correctly identified the mismatch (label: 0), as the code converts objects to JSON strings rather than converting Python strings to C-style strings. This demonstrates how human annotators can be misled by superficial keyword matching while agents focus on actual functionality.}

\revised{\textbf{Case 6}: For query "python print json tree," human annotators incorrectly labeled a JSON export function as matching (label: 1). The test-driven agent correctly identified the mismatch (label: 0), as the code returns JSON data without printing it.}

\revised{\subsection{Semantic Misalignment Cases}}
\revised{\textbf{Case 7}: Query "how to get encoding type in python" was misaligned with file encoding detection code. The agent failed to understand that "encoding type" in this context specifically refers to file character encoding rather than general encoding concepts, leading to incorrect test generation that didn't match the file-specific functionality.}

\revised{\textbf{Case 8}: Query "angle between two vectors using python" matched with code that calculates both single vector angles and inter-vector angles. The agent struggled to capture the precise behavioral intent, generating tests for one interpretation while missing the dual-mode functionality that computes different angle types based on argument presence.}

\revised{\subsection{Validation Misalignment Cases}}
\revised{\textbf{Case 9}: The test program validates the code's method functionality but fails to verify if it actually addresses the query's requirement. The query asks for capturing standard output stream "into a variable in local namespace". However, the code is a class method that retrieves cached output from different sources, not a solution for capturing live stdout into local namespace variables. The test program incorrectly validates the method's output retrieval capabilities rather than testing whether it solves the user's actual problem of stdout capture in local scope.}

\revised{\textbf{Case 10}: The provided code included additional comma-separated string parsing logic beyond the query scope. The test program validated all code functionalities instead of focusing solely on array-to-list conversion, leading to false positive labeling.}

\revised{These cases underscore that while our test-driven approach significantly improves annotation accuracy, challenges remain in handling natural language ambiguity and ensuring test programs align precisely with query intent rather than code functionality.}

\begin{table*}[!ht]
\centering
\caption{Failure Examples of Query-Code Pairs}
\label{tab:query_code_examples}
\normalsize  
\renewcommand{\arraystretch}{1.2}
\begin{tabular}{|>{\centering}m{0.05\textwidth}|>{\raggedright\arraybackslash}m{0.22\textwidth}|>{\raggedright\arraybackslash}m{0.65\textwidth}|}
\hline
\centering\textbf{Case} & \centering\textbf{Query} & \centering\textbf{Code}\arraybackslash \\[3pt]
\hline
1 & python fabric verbose logging & 
\begin{lstlisting}[language=Python, basicstyle=\small\ttfamily, breaklines=true]
def info(self, *args, **kwargs):
    return self._write(args, level=INFO, 
                      console=kwargs.get("console", False))
\end{lstlisting} \\
\hline
2 & slice every 5 items python & 
\begin{lstlisting}[language=Python, basicstyle=\small\ttfamily, breaklines=true]
def _(xs: Iterable, n: int)->Iterable:
    return IterableIterMap(xs, itertools.islice(0, n))
\end{lstlisting} \\
\hline
3 & lowercase + string object + python & 
\begin{lstlisting}[language=Python, basicstyle=\small\ttfamily, breaklines=true]
def to_string(value):
    if isinstance(value, str):
        return value
    elif isinstance(value, bool):
        return str(value).lower()
    else:
        return str(value)
\end{lstlisting} \\
\hline
4 & how to get next month python datetime & 
\begin{lstlisting}[language=Python, basicstyle=\small\ttfamily, breaklines=true]
def get_next(self, ret_type=datetime):
    self.time = self.time + timedelta(seconds=self.interval)
    if ret_type == datetime:
        return self.time
    elif ret_type == float:
        return time.mktime(self.time.timetuple())
\end{lstlisting} \\
\hline
5 & python string to cstr & 
\begin{lstlisting}[language=Python, basicstyle=\small\ttfamily, breaklines=true]
def string(value) -> str: 
    """ string dict/object/value to JSON """ 
    return system_json.dumps(Json(value).safe_object(), ensure_ascii=False)
\end{lstlisting} \\
\hline
6 & python print json tree & 
\begin{lstlisting}[language=Python, basicstyle=\small\ttfamily, breaklines=true]
def export(self, node):
    """Return JSON for tree starting at `node`."""
    dictexporter = self.dictexporter or DictExporter()
    data = dictexporter.export(node)
    return json.dumps(data, **self.kwargs)
\end{lstlisting} \\
\hline
7 & how to get encoding type in python & 
\begin{lstlisting}[language=Python, basicstyle=\small\ttfamily, breaklines=true]
def _get_file_encoding(self, filename):
    f = self._get_or_open_file(filename)
    try:
        encoding = f.encoding
    except AttributeError as e:
        # If f doesn't have an encoding attribute,
        # raise a TypeError
        if e.args == ('encoding',):
            raise TypeError('Expected file-like object')
        raise e  # pragma: no cover
    finally:
        f.close()
    return encoding
\end{lstlisting} \\
\hline
\end{tabular}
\end{table*}

\begin{table*}[!ht]
\centering
\normalsize  
\renewcommand{\arraystretch}{1.2}
\begin{tabular}{|>{\centering}m{0.05\textwidth}|>{\raggedright\arraybackslash}m{0.22\textwidth}|>{\raggedright\arraybackslash}m{0.65\textwidth}|}
\hline
\centering\textbf{Case} & \centering\textbf{Query} & \centering\textbf{Code}\arraybackslash \\[3pt]
\hline
8 & angle between two vectors using python & 
\begin{lstlisting}[language=Python, basicstyle=\small\ttfamily, breaklines=true]
def angle(self, *args):
    x = self.dx
    y = self.dy
    z = self.dz
    
    if len(args) == 0:
        if self.mag() < res:
            return 0.0
        if x >= 0 and y >= 0:
            try:
                return math.atan(y / x)
            except ZeroDivisionError:
                return math.pi / 2
        elif x < 0 and y >= 0:
            return math.pi - math.atan(y / abs(x))
        elif x >= 0 and y < 0:
            try:
                return -math.atan(abs(y) / x)
            except ZeroDivisionError:
                return -math.pi / 2
        else:
            return math.atan(abs(y) / abs(x)) - math.pi
    elif len(args) == 1:
        b = args[0]
        try:
            rv = math.acos(self.dot(b) / (self.mag() * b.mag()))
            return rv
        except ZeroDivisionError:
            return 0.0
\end{lstlisting} \\
\hline
9 & capture the standard output stream of a cell into a variable in local namespace python & 
\begin{lstlisting}[language=Python, basicstyle=\small\ttfamily, breaklines=true]
def out(self):
    """Std/out output (cached)"""
    if self.__out is not None:
        return self.__out
    if self._uses_subprocess:
        self.__out = self.std_out.read()
    else:
        self.__out = self._pexpect_out
    return self.__out
\end{lstlisting} \\
\hline
10 & how to change array of strings to list of strings in python & 
\begin{lstlisting}[language=Python, basicstyle=\small\ttfamily, breaklines=true]
def getlist(v):
    if v is None:
        return None
    if isinstance(v, str):
        v = v.split(",")
    elif not hasattr(v, "__iter__"):
        v = [v]
    return list(v)
\end{lstlisting} \\
\hline
\end{tabular}
\end{table*}

\afterpage{\clearpage
\vspace*{-2\baselineskip}
\MRrevised{\section{Case Study: The Role of the Final Arbiter}}

\MRrevised{This case study illustrates a scenario where a code snippet passes the generated tests but is correctly rejected by the Final Arbiter, demonstrating its role in preventing false positives through semantic analysis beyond test execution.}

\MRrevised{As shown in Figure \ref{test_pass_rejected}, the test program for the query \texttt{"python get timestamp seconds since epoch"} passed. However, the Final Arbiter identified a fundamental mismatch: the query requires retrieving the \textit{current} timestamp, while the provided function only converts an existing ISO 8601 string. Therefore, the arbiter correctly rejected the code, highlighting its critical role in verifying the query's intent rather than just the test result, thus enhancing the robustness of our retrieval system.}

\begin{figure}
    \centering
    \includegraphics[width=1.0\linewidth]{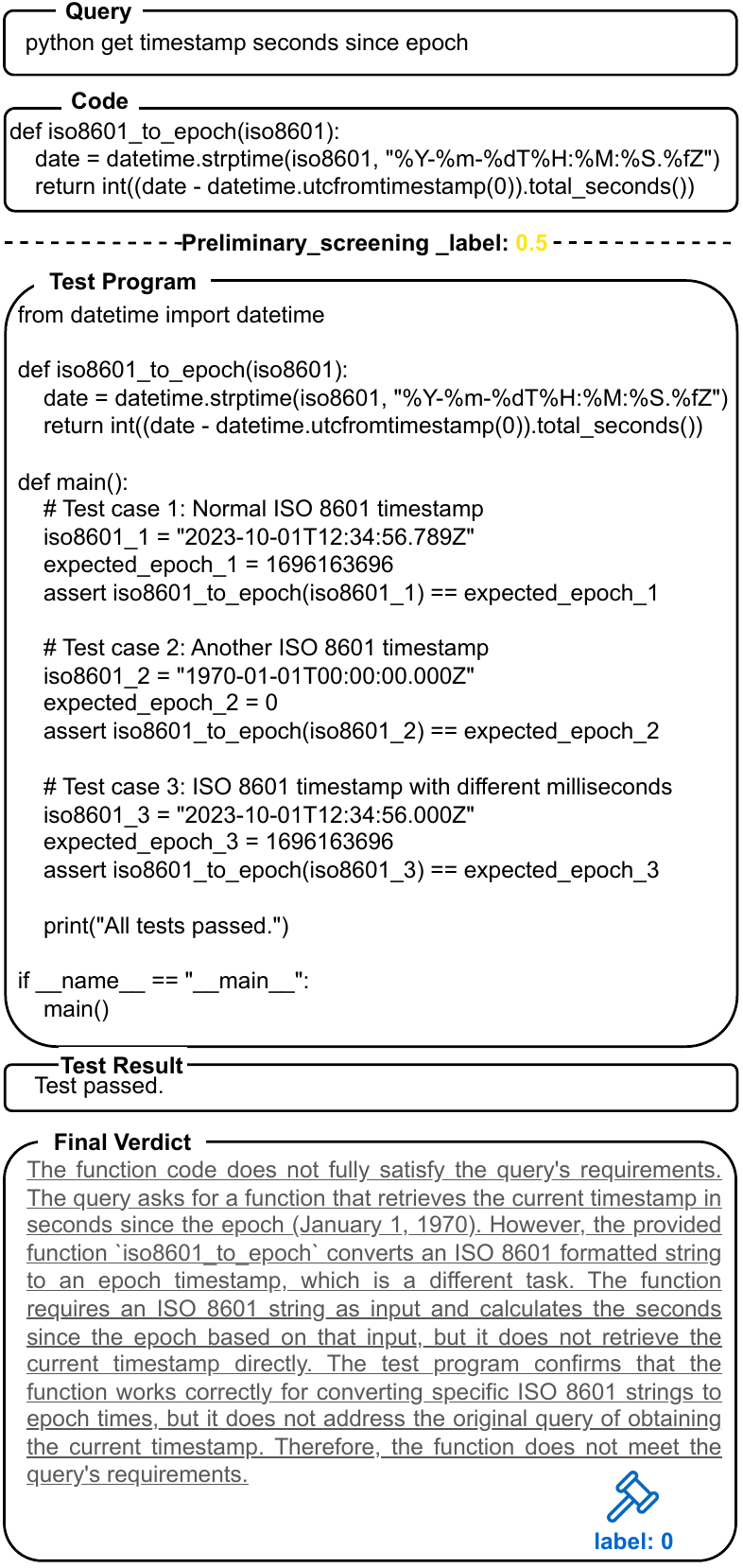}
    \caption{\MRrevised{An example where a code snippet passes the generated tests but is correctly rejected by the final arbiter for failing to meet the requirement of the query.}}
    \label{test_pass_rejected}
\end{figure}
}

\end{document}